\documentclass[aps,prb,citeautoscript,reprint,superscriptaddress,amsmath,amssymb]{revtex4-2}

\usepackage{graphicx}% Include figure files
\usepackage{dcolumn}% Align table columns on decimal point
\usepackage{bm}% bold math
\usepackage{hyperref}% add hypertext capabilities
\usepackage{cancel}

\usepackage{multirow} %Cosa para tablas
\usepackage{graphicx}
 \usepackage{float} %figure inside minipage

\usepackage[version=4]{mhchem}
\usepackage[normalem]{ulem}
\usepackage{xcolor}

\begin{document}

\title{Optimizing Hubbard U parameters for Enhanced Description of Electronic and Magnetic Properties in CrI$_3$ Monolayers and Bilayers}

\author{Diego Lauer}
\affiliation{Departamento de F\'{i}sica, Universidad 
T\'{e}cnica Federico Santa Mar\'{i}a, Casilla Postal 
110V, Valpara\'{i}so, Chile.}
\affiliation{Centro de Física de Materiales-Materials Physics Center CFM-MPC CSIC-UPV/EHU, Donostia International Physics Center, Paseo Manuel Lardizabal 5, Donostia–San Sebastián 20018, Spain}

\author{Jhon W. Gonz\'{a}lez}
\thanks{Corresponding author}
\email{jhon.gonzalez@uantof.cl}
\affiliation{Departamento de Física, Universidad de Antofagasta, Avenida Angamos 601, Casilla 170, Antofagasta, Chile}

\author{Eric Su\'{a}rez Morell}
\affiliation{Departamento de F\'{i}sica, Universidad 
T\'{e}cnica Federico Santa Mar\'{i}a, Casilla Postal 
110V, Valpara\'{i}so, Chile.}

\author{Andr\'{e}s Ayuela}
\thanks{Corresponding author}
\email{a.ayuela@csic.es}
\affiliation{Centro de Física de Materiales-Materials Physics Center CFM-MPC CSIC-UPV/EHU, Donostia International Physics Center, Paseo Manuel Lardizabal 5, Donostia–San Sebastián 20018, Spain}

\date{\today}% It is always \today

\begin{abstract}
The magnetic properties of CrI$_3$ monolayers, which were recently measured, have been investigated considering electronic repulsion and localization effects in Cr 3d orbitals. In this study, we propose a DFT approach using Hubbard U corrections to improve accuracy.  We compare the valence density-of-states using the HSE06 hybrid functional and the DFT+U approach, which includes U parameters for both Cr 3d and I 5p orbitals. The results of our study indicate that the optimal values for U(Cr$_{3d}$) and U(I$_{5p}$) are 3.5 eV and 2.0 eV, respectively.   This approach is further applied to improve calculations of electronic and magnetic properties in CrI$_3$ monolayers and, more importantly, in bilayers combined with van der Waals functionals.  These refinements hold promise for further studies of complex CrI$_3$ nanostructures, and may also be of interest for other trihalide few-layer systems.
\end{abstract}

\maketitle

\section{\label{sec:intro}Introduction}
\begin{figure}[b!]
\centering
\includegraphics[clip,width=0.99\columnwidth]{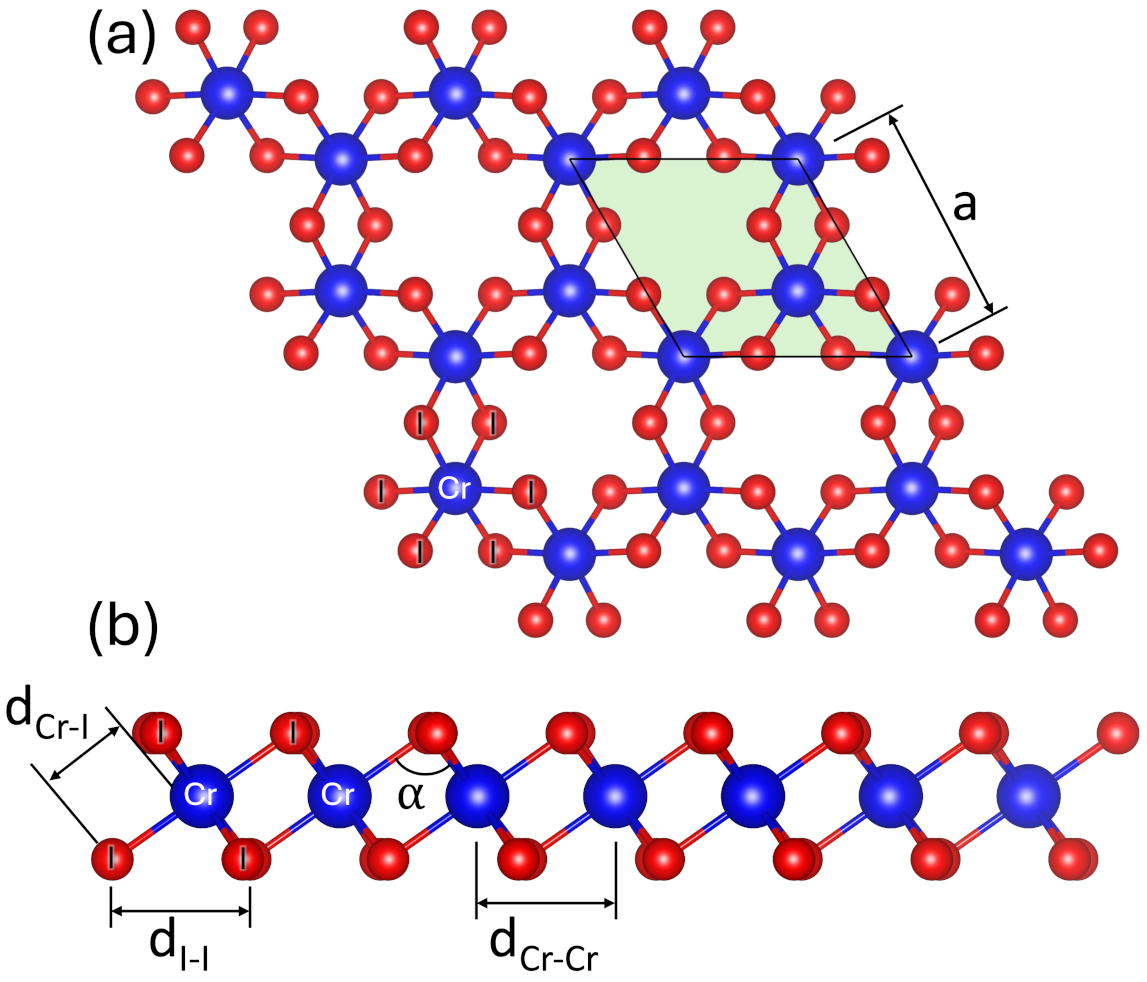}
    \caption{Schematic of the atomic structure of a CrI$_3$ monolayer. (a) The top view along the c-direction shows the hexagonal lattice with the unit cell highlighted in green and the lattice constant $a$. (b) Side view along the a-direction, indicating key structural parameters: the Cr–I bond length ($d_{\text{Cr–I}}$), I–I distance ($d_{\text{I–I}}$), Cr–Cr distance ($d_{\text{Cr–Cr}}$), and the Cr–I–Cr bond angle ($\alpha$). Chromium and iodine atoms are depicted in blue and red, respectively.}
    \label{fig:CrI3-estructura}
\end{figure}

Chromium trihalides (\ce{CrX3}) are part of the van der Waals (vdW) layered material family. These materials have a hexagonal in-plane structure and are held together out-of-plane by weak vdW interactions. In particular, monolayer CrI$_3$—first synthesized in 2017—was among the first intrinsic magnetic two-dimensional materials \cite{huang2017layer}. It consists of two chromium atoms and six iodine atoms per unit cell (Figure~\ref{fig:CrI3-estructura}), and displays ferromagnetic behavior with a magnetic moment of $3.0\, \mu$B/Cr \cite{liu2016exfoliating}, a tunable Curie temperature of 45 K \cite{huang2017layer,yang2021enhancing}, and a band gap of 1.2 eV \cite{wang2011electronic,zhang2015robust,mcguire2015coupling}. Magnetic interactions in the monolayer are mediated not directly between Cr atoms but through I atoms via a superexchange mechanism, which also contributes to its out-of-plane magnetic anisotropy \cite{lado2017origin}.

Despite advances in density functional theory (DFT), accurately modeling trihalides like CrI$_3$ remains challenging due to self-interaction errors that impact  mainly on the localization of 3d electrons. Two widely adopted strategies to address this issue are hybrid functionals \cite{garza2016predicting} and DFT+U corrections \cite{dudarev1998electron}. Hybrid functionals such as HSE06 \cite{krukau2006influence} incorporate a portion of exact exchange, improving predictions of band gaps, electronic structures, and density of states in a range of materials \cite{ibragimova2018first,yang2017understanding,schafer2021cerium,li2020assessment}. However, their high computational cost—typically over 70\% more than GGA-based methods \cite{pandey2017electronic}—and today incompatibility with non-local vdW functionals \cite{dion2004van,roman2009efficient,klimevs2011van} limit their scalability for large or defected systems.

In contrast, the DFT+U method provides a computationally efficient correction by introducing an on-site Coulomb repulsion term. It has been widely used to model CrI$_3$ monolayers, with reported U values for the Cr 3d orbitals ranging from 1 to 6 eV \cite{yang2021enhancing,kashin2020orbitally,haddadi2023site,kumar2019magnetism}. Applying U to these orbitals improves electron localization, lowers the spin-minority conduction bands by up to 3 eV \cite{haddadi2023site}, and changing the character of top valence band from Cr-3d to I-5p  \cite{kundu2020valence}.

Optimal U values are often calibrated by matching selected properties—such as band gaps or lattice constants to experiments or hybrid functional benchmarks \cite{hong2012spin}. While DFT+U tends to underestimate band gaps and HSE06 tends to overestimate them, these methods effectively bracket experimental values \cite{li2013density}, with DFT+U offering comparable accuracy at significantly lower computational costs \cite{pandey2017electronic}. 

Though originally developed for d- and f-electrons, recent studies have shown that U corrections applied to p orbitals of non-metal atoms can further improve agreement with experiments and hybrid functionals. In metal oxides such as \ce{Fe2O3}, NiO, \ce{TiO2}, \ce{ZrO2}, and \ce{HfO2}, applying U to both 3d and 2p orbitals improves formation energies and phase diagrams \cite{pandey2017electronic,huang2016surface,huang2016surface}.

In this work, we present a refined method to calibrate Hubbard U parameters for CrI$_3$ layers by aligning DFT+U densities of states with those obtained using the HSE06 hybrid functional. We show that the best agreement is achieved by including U corrections on the Cr 3d and I 5p orbitals. These optimized values are then applied to investigate the magnetic ordering in CrI$_3$ monolayers and bilayers, with particular attention to different stacking configurations. Our findings improve the description of interlayer coupling and are valuable for future studies on few-layer CrI$_3$ and related nanostructures.

Although our primary focus is on CrI$_3$, our proposed methodology broadly applies to other magnetic layered systems with strong electronic correlations. This approach provides a transferable and efficient framework for studying electronic and magnetic properties in novel two-dimensional materials for spintronic applications.

\section{\label{sec:Method}Methodology and computational details}

Our systems are modeled in the framework of density functional theory using the projector augmented wave (PAW) method implemented in the Vienna ab initio simulation package (VASP) \cite{kresse1996efficient}. We use a PBE-GGA approach to represent the exchange and correlation interactions \cite{perdew1996generalized}. To correct the electronic localization problems, which mainly affect the d-electrons, we include a Hubbard correction term (DFT+U) using the Dudarev rotationally invariant approximation \cite{dudarev1998electron}.

 Our DFT+U calculations are compared with the calculations of the Heyd-Scuseria-Ernzerhof exchange correlation functional HSE06 screened \cite{krukau2006influence} because of the better prediction of electronic properties by hybrid functionals \cite{garza2016predicting}. The HSE06 functional provides a more accurate representation of the layer band structure. However, it is a computationally demanding method and there are technical limitations, such as the incompatibility of including many-body corrections for van der Waals interactions of interest in 2D materials for studying stackings.
In our DFT+U calculations of CrI$_3$ layers, we consider three parameter settings: without any Hubbard correction ($U_{Cr,I}=\left\lbrace 0.0\, ,0.0 \right\rbrace$), the traditional approach with Hubbard-U correction only in the d-orbitals of the chromium atom ($U_{Cr,I}=\left\lbrace U_d\,,0.0 \right\rbrace$), and a novel approach including the repulsion Hubbard-U term to the d-orbitals of the chromium atom and p-orbitals of the iodine atom ($U_{Cr,I}=\left\lbrace U_d\,,U_p \right\rbrace$). We relax the geometry of the layer for each value of Hubbard U parameter\cite{huang2016surface}.

After performing the convergence tests and considering a tolerance of less than $10^{-5}$ eV/atom, we use a $\Gamma$ centered Monkhorst-Pack k-mesh of $6 \times 6 \times 1$ for structural relaxation and a denser $\Gamma$ centered k-mesh of $15 \times 15 \times 1$ to calculate the density of states (DOS). We set an energy cutoff of $550$ eV for plane waves, a total energy convergence criterion of 10$^{-8}$ eV, and a force convergence criterion of 10$^{-3}$ eV/\AA.
A vacuum region of $20$ \AA{} in the out-of-plane direction is included to reduce the interaction between images. We used the VASPKIT code to post-process the calculated data \cite{VASPKIT}. 

%%%%
We used the average Pearson correlation coefficient $\mathcal{P}$ to find the optimal U values by comparing the densities of states of the DFT+U calculations with those of HSE06 \cite{wooldridge2006introduccion,BERMAN2016135,howell2010log}.
The average Pearson correlation coefficient is defined as,  
\begin{equation}
\mathcal{P}_i=\frac{N\left(\sum \mathcal{L}_U\,\mathcal{L}_H-\sum \mathcal{L}_U \sum \mathcal{L}_H\right)}{\sqrt{\left(N\sum \mathcal{L}_U^2-\left(\sum \mathcal{L}_U\right)^2\right)\left(N\sum \mathcal{L}_H^2-\left(\sum \mathcal{L}_H\right)^2\right)}}, \nonumber
\end{equation}
\begin{equation}
\mathcal{P}= \frac{\mathcal{P}_{\uparrow}+\mathcal{P}_{\downarrow}}{2},
\end{equation}
where $\mathcal{P}_i$ is the Pearson correlation coefficient for spin-component $i$  ($i=\uparrow, \downarrow$), $N$ is the number of data points and $\mathcal{L}_U,\mathcal{L}_H$ are the density of states from DFT+\textit{U} and hybrid HSE06 functional, respectively.

Our calculations show that the density of states calculated using hybrid functionals is more extended in energy than that calculated with DFT+U. In contrast, the one obtained using DFT+U is compressed in energy with respect to HSE06, in agreement with previous works \cite{ibragimova2018first,nisar2011optical,yang2017understanding}.
To compare the DOSs directly, we fit a scaling factor that adjusts the energy of the DFT+U DOS with the results of the hybrid functional HSE06. The energy scaling parameter, $\varepsilon$, is used to align the density of states in a particular range of energies by multiplying the DFT+U energy axis.
To find the optimal value of $\varepsilon$, we multiply the energy of the DOS (with $E_F = 0$ eV) by this positive factor and interpolate the density of states for a one-to-one comparison with the HSE06 density. We then calculated the average Pearson correlation between the scaled DFT+U and HSE06 for each scaling factor. The optimal value of $\varepsilon$ corresponds to the highest correlation average for spin-up and spin-down states.
 
By performing this correlation analysis, we systematically seek the maximum average correlation coefficient value, which implies that we have a better agreement between the DFT+U  and hybrid HSE06 calculations.
Our analysis includes three cases: DFT without Hubbard-U corrections, DFT with Hubbard-U correction only in the 3d levels of the chromium atom, and DFT with Hubbard-U correction in the 2p levels of the iodine atom and 3d levels of the chromium atom. This methodology allows us to identify the case that best matches the hybrid functional electronic density of states.

We demonstrate the advantages of this approach for the CrI$_3$ monolayer by calculating the magnetic order and the anisotropy energy (MAE), which is given by $MAE=E_{in-plane}-E_{out-plane}$. We also investigated the magnetic order of the CrI$_3$ bilayer in the high-temperature (HT) and low-temperature (LT) crystallographic phases observed in experiments.

\section{\label{sec:Analysis and Results} Results and Discussions}
To find the optimal values, we focus on the density of states between -6.0 eV and 0.0 eV. We scan the energy scaling $\varepsilon$ in the range [0.5, 1.5] 
and consider a parameter space for $U_{Cr,I}=\left\lbrace U_d\,,U_p \right\rbrace$, where $U_d$ changes in the range [0.0, 7.0] eV for d-orbitals of the Cr atom, and $U_p$ changes in the range [0.0, 5.0] eV for p-orbitals of the I atom. 

The Pearson correlation value $\mathcal{P}$ obtained for the three parameter sets considered is summarized in Table \ref{tab:Tabla-Pearson-Correlation-Results}. The first with $U_d=U_p=0$ and then the two best results, varying first, only $U_d$, and then with a fixed $U_d$, we obtain the best $U_p$. The results for the entire set of parameters used can be seen in Figure S1 of the Supplementary Material. 
Figure S1 shows a systematic analysis of the Pearson correlation coefficient in various Hubbard parameter settings, supporting our choice of optimal parameters such as $U_{Cr, I}=\left\lbrace 3.5, 2.0 \right\rbrace$, which achieves the best agreement (Pearson correlation of $\mathcal{P}=0.95$).

The correlation without Hubbard ($U_{Cr,I}=\left\lbrace 0\,,0 \right\rbrace$) is the lowest among the alternatives, with a Pearson correlation coefficient of $\mathcal{P}=0.78 $. Introducing Hubbard U to the chromium atom ($U_{Cr,I}=\left\lbrace U_d\,,0 \right\rbrace$), the correlation value increases substantially to $\mathcal{P}=0.93$, indicating a very good agreement. The Hubbard U configuration on the p orbitals of the iodine and the d orbitals of the chromium atoms ($U_{Cr,I}=\left\lbrace U_d\,,U_p \right\rbrace$) is the best option, yielding the highest correlation, with a value of $\mathcal{P}=0.95$ and achieving closer agreement with the electronic properties predicted by the hybrid functional HSE06. The calculations with spin-orbit coupling follow the same trend and perform even better.

Including the Hubbard-U term in both atoms is the best way to represent the DOS accurately while cutting computational costs. Additionally, DFT+U allows us to include many-body corrections to the van der Waals interactions (vdw-DF) necessary to correctly describe magnetism in the bilayers, specifically, in the HT stacking \cite{mcguire2015coupling}. To our knowledge, at the time of this publication, it is not possible to simultaneously include both HSE06 and vdw-DF in the same calculation.

\begin{table}[t]
\begin{center}
\caption{Average Pearson correlation $\mathcal{P} $ and optimal energy scaling $\varepsilon$ values for different Hubbard U configurations with and without spin-orbit effects. For additional details, please refer to Fig. S1.}
\begin{tabular}{c c c c c }
\hline 
\hline 
$U=\left\lbrace U_{Cr_d},U_{I_p}\, \right\rbrace$  & $\mathcal{P} $    & $\varepsilon$ & $\mathcal{P}_{SOC}$  & $\varepsilon_{SOC}$ \\  
\hline 
\hline 
$\left\lbrace 0.0\,,0.0 \right\rbrace$ eV & $0.78$ & $1.08$ & $0.84$ & $1.13$ \\ 

$\left\lbrace 3.5\,,0.0 \right\rbrace$  eV & $0.93$ & $1.13$ & $0.96$ & $1.16$ \\ 

$\left\lbrace 3.5\,,2.0 \right\rbrace$ eV & $0.95$ & $1.13$ & $0.98$ & $1.15$ \\ 
\hline 
\hline 
\end{tabular} 
\label{tab:Tabla-Pearson-Correlation-Results}
\end{center}
\end{table}

In Figure \ref{fig:DOSCrI3-escalado}, we compare the DOS obtained with the three sets of parameters in Table \ref{tab:Tabla-Pearson-Correlation-Results} with that calculated with the hybrid potential HSE06 (shown as a shaded region in all three panels). The DFT+U results are shown as a red dashed line, and the scaled DFT+U DOS calculations are shown as a solid black line.

Using U on Cr improves the lower part of the valence band in the range of -6 to -3 eV by shifting down the Cr states. However, using U on Cr and I is an effective method for adjusting not only the lower part of the valence band but also the middle and top parts, as the hybridization between Cr and I is slightly reduced. 
As discussed in the following, further studies on magnetic properties and stacking order are essential to fully understand the combined effects of the U parameters.

\subsection{Without Spin-Orbit Coupling}
\begin{figure}[ht!]
\includegraphics[clip,width=1\columnwidth,angle=0,clip]{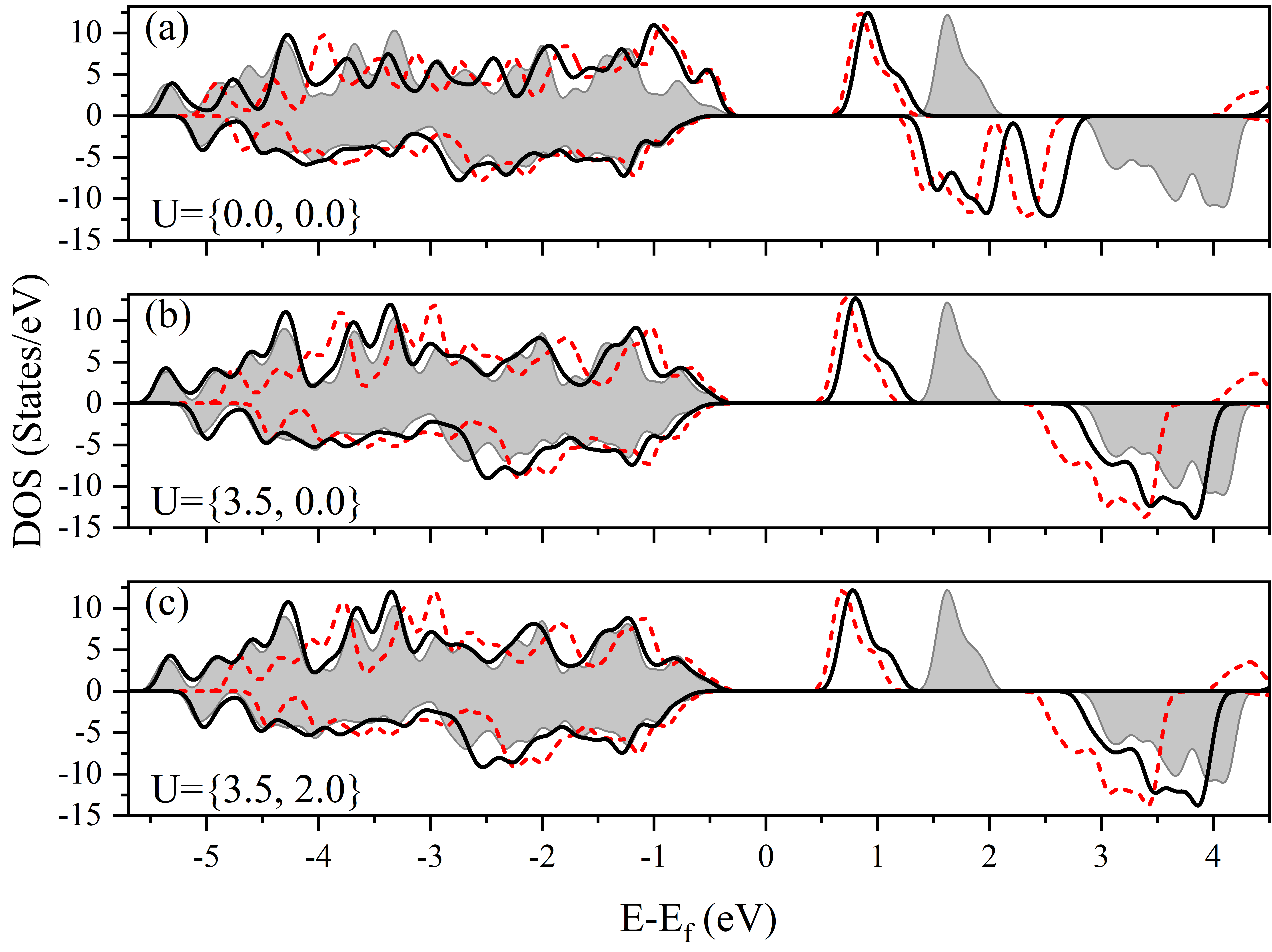}
    \caption{Density of states of CrI$_3$ monolayer within different models. The calculation using the HSE06 hybrid functional is used as a reference and is shown as a shadow region. 
    The raw DFT+U density of states is represented by a dashed red line, and the energy-scaled DFT+U is represented by a black line. 
    DOS results are given using (a) $U_{Cr,I}=\left\lbrace 0.0\,,0.0 \right\rbrace$ ($\mathcal{P}=0.78$),  (b) $U_{Cr,I}=\left\lbrace 3.5\,,0.0 \right\rbrace$, ($\mathcal{P}=0.93$), and (c) $U_{Cr,I}=\left\lbrace 3.5\,,2.0 \right\rbrace$ ($\mathcal{P}=0.95$). }
    \label{fig:DOSCrI3-escalado}
\end{figure}

Figure \ref{fig:DOSCrI3-escalado} clearly illustrates the compression of the density of states (DOS) calculated using DFT+U (red line) compared to the HSE06 reference (shaded region). By incorporating an energy scaling factor (black line), the compression caused by electronic correlation effects is effectively corrected, achieving better agreement with the HSE06 results.
 The band gap calculated with HSE06 is the largest, approximately $1.95$ eV, in line with previous results \cite{jiang2018spin,zhang2018strong}. The states above the Fermi level (between 1.5 and 2.0 eV for HSE06) are dominated by d-up electrons \cite{haddadi2023site,kundu2020valence}. In DFT+U, these states move to lower energies, leading to smaller gaps.
 Modifying those states within our parameter space without mismatch in the bands below the Fermi level was challenging,  even when considering Hubbard-J repulsion terms. 
The bandgap for $U_{Cr,I}=\left\lbrace 0\,,0 \right\rbrace$ is $1.19$ eV; for $U_{Cr,I}=\left\lbrace 3.5\,,0.0 \right\rbrace$, $1.04$ eV; for $U_{Cr,I}=\left\lbrace 3.5\,,2.0 \right\rbrace$, $1.02$ eV.

The up-projection of the valence density of states with $U_{Cr,I}=\left\lbrace 0\,,0 \right\rbrace$ in Figure \ref{fig:DOSCrI3-escalado}(a) is a poor match with those using the hybrid functional. 
Fig. S4 of the supplementary material provides a detailed analysis of the atomic contributions to the density of states, clearly showing how the addition of Hubbard terms on iodine orbitals enhances the alignment between DFT+U and hybrid calculations.
The spin-down component aligns precisely quite well with the hybrid functional calculations, but the top of the valence band of the spin-up has extra d-Cr states close to the gap, as shown in Figure S4. This feature is incompatible with DFT results using hybrids and angle-resolved photoemission spectroscopy (ARPES) experiments\cite{haddadi2023site}. 

The U term has a dominant effect on the d-Cr orbitals, as shown in Figure \ref{fig:DOSCrI3-escalado}(b, c),  and the DOS in the -6.0 eV and 0.0 eV region agrees with the hybrid calculations. Furthermore, the U term in the p-I orbitals has markedly improved the result.

Figure \ref{fig:DOSCrI3-escalado}(c) shows that the Hubbard-U term in the 2p states of the iodine atom shifts DOS to the left. This shift results in a much closer agreement with the results of the hybrid functional, as shown in Table \ref{tab:Tabla-Pearson-Correlation-Results}.

We now investigate the variation of the d-Cr and p-I band centers as a function of the Hubbard U$_{Cr, I}$ parameters. The d-band center is a well-established descriptor in catalysis and electrochemical performance studies used to understand the formation bond and the tendency to surface reactivity \cite{jiao2022descriptors, zhu2022d}. It can also serve as an indirect measure of electron-electron repulsion interactions in transition-metal compounds.
This metric allows us to investigate how variations in the Hubbard U parameter modify the electronic structure, providing an indirect insight into electron-electron interactions within the system.
Here, we compute the d-band center for the chromium atoms and the p-band center for the iodine atoms to show the explicit effects of the Hubbard U correction on the electronic structure.

We calculate the band center using:
\begin{equation}
\varepsilon = \frac{\int_{-\infty}^{E_F} E \, \rho(E)dE}
{\int_{-\infty}^{E_F} \, \rho(E)dE },
\end{equation}
where $\rho(E)$ represents the density of states summed over the orbitals (d-Cr, p-I), and E denotes the energy \cite{demiroglu2016dft}.

\begin{figure}[ht!]
\includegraphics[clip,width=1\columnwidth,angle=0,clip]{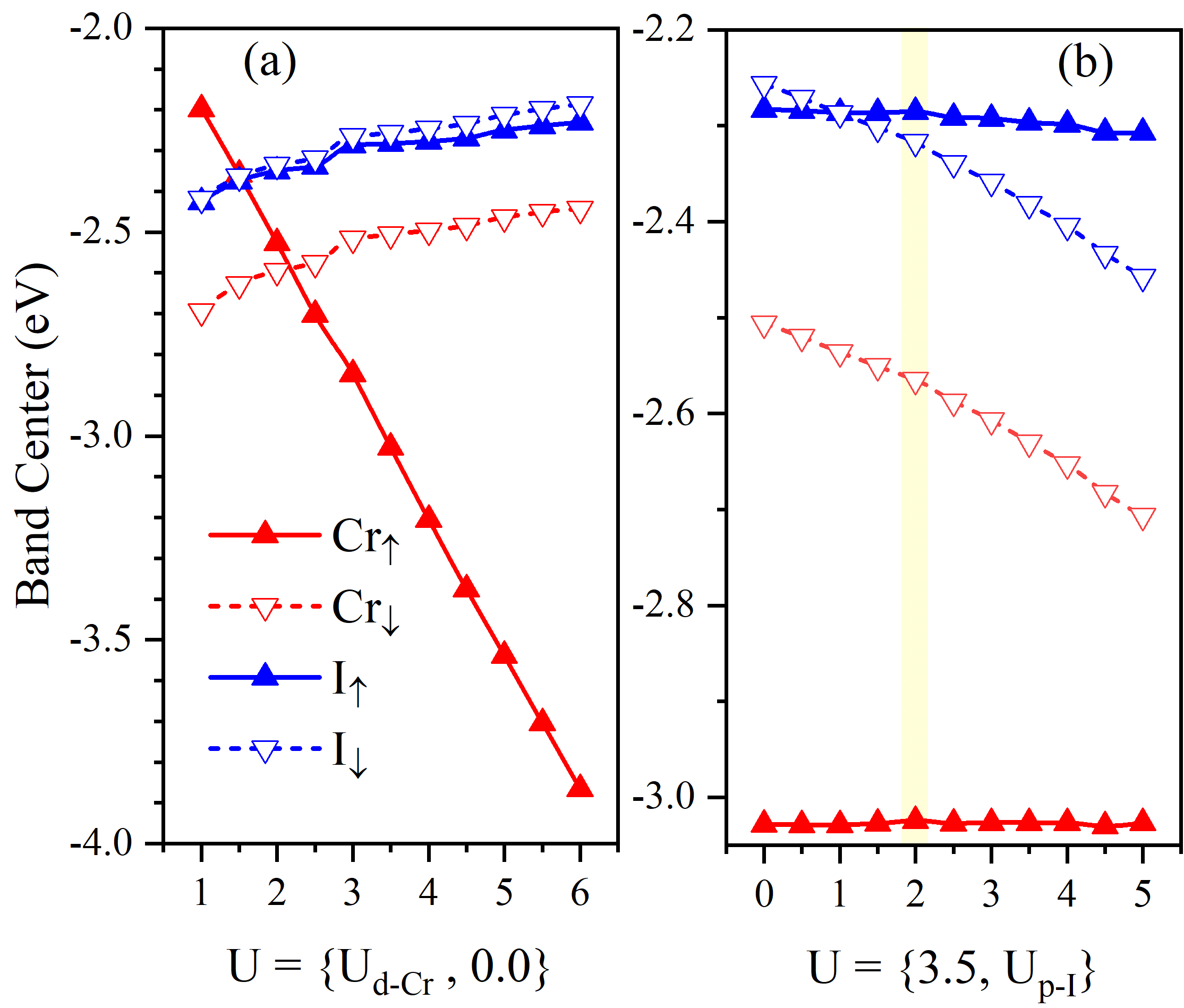}
\caption{CrI$_3$ band center energies as a function of the Hubbard U values on d-Cr and p-I orbitals. The red lines depict the Cr d-band centers in spin-up ($\uparrow$) and spin-down ($\downarrow$) states, while the blue lines represent the I p-band centers for the same spin states. (a) The variation of band centers with respect to $U_{d-Cr}$ in the range [1.0, 6.0] eV while keeping $U_{p-I}$ fixed at 0.0 eV. (b) The variation of band centers as a function of $U_{p-I}$ in the range [3.0, 5.5] eV for $U_{d-Cr}$ fixed at 3.5 eV. The yellow highlighted region in (b) marks the point of best correlation between HSE06-hybrid and DFT+U, corresponding to $U_{d-Cr} = 3.5$ eV and $U_{p-I} = 2.0$ eV.}
\label{fig:Band-CENTER-no-escalado}
\end{figure}

Figure \ref{fig:Band-CENTER-no-escalado}(a) shows the variation of $U_{d-Cr}$ with $U_{p-I}$ set to zero. As U is applied to the d-Cr orbitals, the Cr$_\uparrow$ d-band center shifts towards more negative values showing a linear trend going from -2.2 eV for $U_{Cr, I}=\left\lbrace 0\,,0 \right\rbrace$ to -3.9 eV for $U_{Cr, I}=\left\lbrace 6\,,0 \right\rbrace$ eV, while the Cr$_\downarrow$ d-band and both I-p band centers follow a similar trend, moving up the band centers with a variation of $\sim 0.3$ eV.

In Figure \ref{fig:Band-CENTER-no-escalado}(b) when the $U_{d-Cr}$ is fixed at 3.5 eV ($U_{Cr,I}=\left\lbrace 3.5\,,U_{p-I} \right\rbrace$ eV).  The Cr$_\uparrow$ d-band and I$_\uparrow$ p-band centers present small variations. However, the Cr$_\downarrow$ d-band and I$_\downarrow$ p-band centers decrease in $\sim 0.2$ eV as we increase the $U_{p-I}$ value, a variation significant enough to be observable.
This behavior aligns with the spin-down density of states presented in Fig. S4, where iodine atoms make a dominant contribution, leading to more significant changes in the spin-down states compared to the spin-up states.

The sensitivity of the band centers to $U_{d-Cr}$ and $U_{p-I}$ points out the high correlation and interaction between the chromium and iodine orbitals, indicating the need to include both parameters for a more accurate representation of the electronic structure and magnetic properties.

\subsection{With Spin-Orbit Coupling}
\begin{figure}
\includegraphics[clip,width=0.49\textwidth,angle=0,clip]{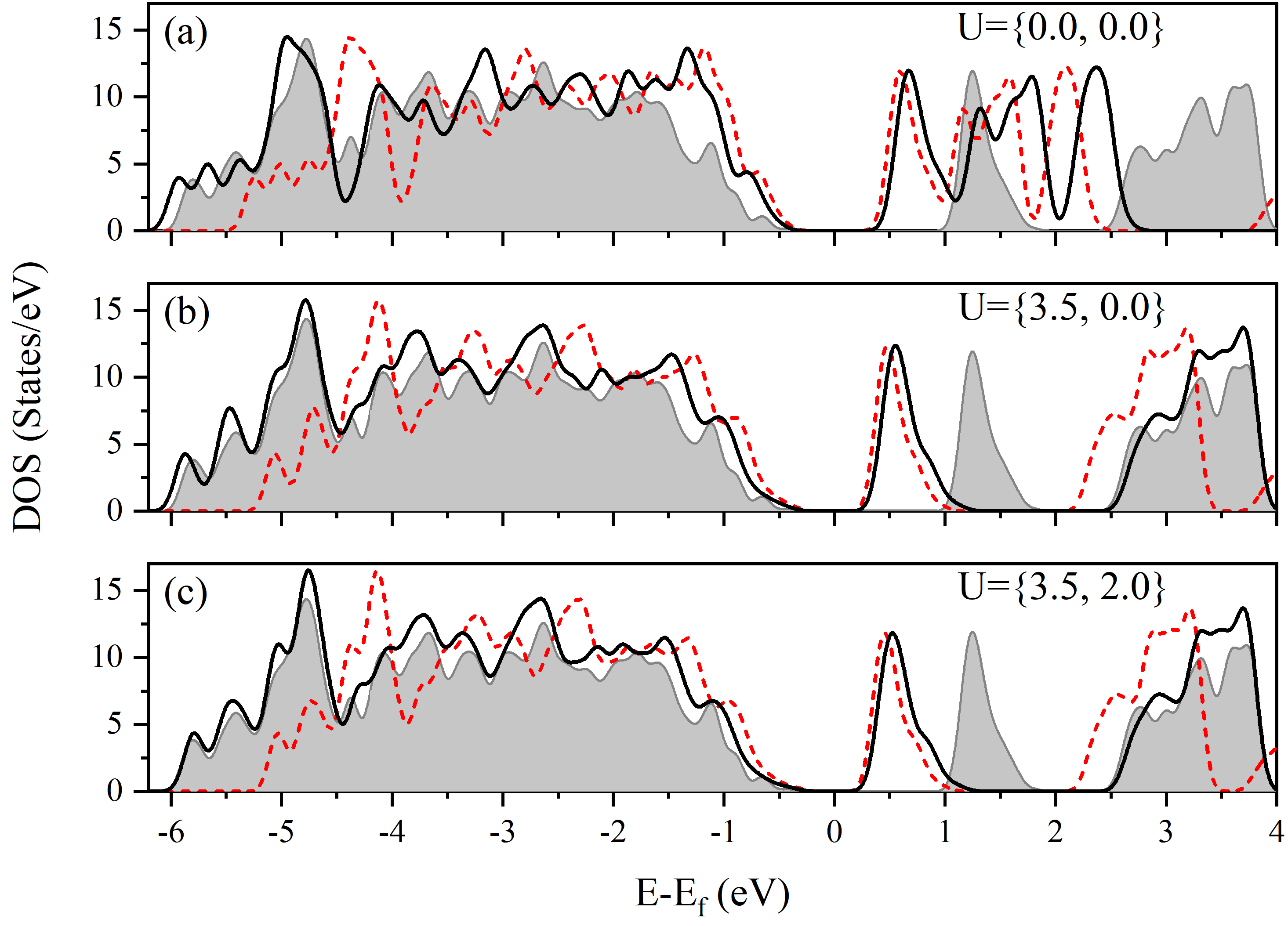}
    \caption{Density of states of CrI$_3$ with different settings of Hubbard parameters including spin-orbit coupling. The color scale and the plot distribution is similar to Fig. \ref{fig:DOSCrI3-escalado}.
    In panel (a), results for $U_{Cr,I}=\left\lbrace 0.0\,,0.0 \right\rbrace$ ($\mathcal{P}=0.84$), in (b) $U_{Cr,I}=\left\lbrace 3.5\,,0.0 \right\rbrace$, ($\mathcal{P}=0.96$), and (c) $U_{Cr,I}=\left\lbrace 3.5\,,2.0 \right\rbrace$ ($\mathcal{P}=0.98$).}
    \label{fig:DOSCrI3-escalado-SOC}
\end{figure}

Table \ref{tab:Tabla-Pearson-Correlation-Results} shows how the correlation $\mathcal{P}$ between DFT+U and hybrid potentials improves when considering the effects of spin-orbit coupling. The up and down magnetic projections are mixed due to the spin-orbit term. The DOS including spin-orbit coupling for the different U settings, along with their projections onto atomic species, is shown in Fig. S2. The GGA DOS without Hubbard correction in Figure \ref{fig:DOSCrI3-escalado-SOC}(a) has the worst correlation with hybrid calculations with a Pearson value, $\mathcal{P} = 0.84$. The GGA (black line) shows the d-states of the chromium atom near the Fermi level (between -2.0 and 0.0 eV), which are absent in HSE06 (shaded area).

Comparing Fig. \ref{fig:DOSCrI3-escalado-SOC}(b) and (c), we observe significant improvements when introducing Hubbard corrections. Figure \ref{fig:DOSCrI3-escalado-SOC} (b), with corrections applied only to chromium orbitals ($U_{Cr, I}=\left\lbrace 3.5,0.0 \right\rbrace$), already shows a marked improvement in alignment with the HSE06 reference DOS, achieving a Pearson correlation of $\mathcal{P}=0.96$. However, including Hubbard corrections for the chromium and iodine orbitals ($U_{Cr, I}=\left\lbrace 3.5,2.0 \right\rbrace$), as shown in Fig. \ref{fig:DOSCrI3-escalado-SOC}(c), provides the best match to the hybrid functional, achieving the highest correlation value $\mathcal{P}=0.98$. This result shows how necessary it is to properly account for electron-electron interactions in chromium and iodine orbitals to capture the electronic structure predicted by hybrid functional calculations.

\subsection{Application of the $U_{Cr,I}$ parameters to the physical properties of CrI$_3$ }

 %We have accurately reproduced the DOS of the hybrid functional at a fraction of the computational cost by including a Coulomb U parameter in the localization of the d orbitals of chromium and the p orbitals of iodine.
 
 We now study how these settings in the Hubbard term impact different physical properties of CrI$_3$ layers and compare them with previous studies, starting with the CrI$_3$ monolayer. Table \ref{tab:ALL2} summarizes the physical properties of the CrI$_3$ monolayer calculated with HSE06 and DFT+U.

\begin{table*}
\begin{center}
\caption{Lattice constant (a), distances (Cr-Cr, Cr-I, I-I), and angles ($\measuredangle$ Cr-I-Cr, $\measuredangle$ I-Cr-I) band gap (gap $\uparrow$, $\downarrow$), antiferromagnetic to ferromagnetic (AF-FM) energy (positive value means ferromagnetic configuration), Curie Temperature (T$_C$), magnetic moments (individual values for chromium and iodine, as well as the total magnetic moment),  under various Hubbard U configurations.}
\begin{tabular}{ c    c    c    c    c    c    c    c    c    c    c    c   c   c } 
\hline \hline
$U_{Cr_d,I_p}$ &   a  &d$_{Cr-Cr}$  &  d$_{Cr-I}$  &  d$_{I-I}$  &  $\measuredangle$Cr-I-Cr  &  $\measuredangle$I-Cr-I & AF-FM &  T$_\text{C}$ & m$_{Cr}$ & m$_{I}$  &  m$_{T}$  &   gap$\uparrow$  &  gap$\downarrow$    \\  
(\AA{}) &(\AA{}) & (\AA{}) & (\AA{}) & (Deg) & (Deg) & (eV) &  (meV/Cr) & $(K)$ &  $(\mu_{B})$ & $(\mu_{B})$ &  ($\mu_{B}$)&(eV)&(eV)  \\
\hline \hline
\bf{HSE06} &6.84 & 3.95 & 2.72 & 3.86 & 93.2 & 90.2&  23 & 40 & 3.23 & -0.11 & 5.8 & 1.95 & 3.69 \\                                                                                         
$\left\lbrace 0.0\,,0.0 \right\rbrace$ &6.84 & 3.95 & 2.73 & 3.85 & 93.2 & 90.2&  18 & 30 & 3.02 & -0.69 & 6.0 & 1.19 & 1.93  \\ 
$\left\lbrace 3.5\,,0.0 \right\rbrace $&6.92 & 3.99 & 2.76 & 3.95& 92.8 & 89.3&  27 & 45 & 3.39 & -0.14 & 6.0 & 1.04 & 3.05 \\   
$\mathbf{\left\lbrace 3.5\,,2.0 \right\rbrace }$&6.93 & 4.00 & 2.79 & 3.91 & 92.7 & 88.9&  28 & 48 & 3.43 & -0.17 & 6.0 & 1.02 & 3.13  \\
\hline \hline
\end{tabular}
\label{tab:ALL2}
\end{center}
\end{table*}

\subsubsection{Geometry and Magnetic order}
Including the Hubbard U increases the lattice constant and slightly changes the atomic positions. The HSE06 calculations are performed with the system geometry relaxed with U = 0. 
The magnitude of our lattice constant is approximately $a\sim 6.9$ \AA{}, which is in close agreement with the values previously reported about $7.0$ \AA{} \cite{zhang2015robust,guo2018half,wang2016doping,webster2018strain,mcguire2015coupling}. 
The distance between Cr-I atoms is estimated to be $d_{Cr-I}\sim 2.7$ \AA{}, consistent with previous findings \cite{yang2021enhancing}. When considering the role of U in the I$_p$ orbitals, we find a slight expansion in the Cr-I distance, accompanied by a reduction in the I-I distance. The measured angles show variations of less than 1 $\%$.

The magnetic character of the CrI$_3$ monolayer is determined to remain ferromagnetic in all scenarios, which is consistent with experimental findings.  With the hybrid functional, we obtain an $AF-FM$ energy difference of 23 meV/Cr, similar to previous reports in the literature \cite{ghosh2019structural, guo2018half, wang2016doping}. The DFT calculations, which do not include the Hubbard term ($U_{Cr,I}=\left\lbrace 0.0,0.0 \right\rbrace$), yield a lower result (18 meV/Cr). The calculation using a Hubbard U parameter only on Cr ($U_{Cr,I}=\left\lbrace 3.5,0.0 \right\rbrace$) reinforces the ferromagnetic character and results in an AF-FM energy of 27 meV/Cr, which surpasses that obtained by HSE06. Furthermore, including the U-term on p-orbitals of I ($U_{Cr,I}=\left\lbrace 3.5,2.0 \right\rbrace $) results in an AF-FM energy difference of 28 meV/Cr.

The experimental Curie temperature (T$_C$) of the CrI$_3$ monolayer has been reported as 45 K \cite{huang2017layer}. However, the HSE06 functional predicts a lower value of T$_C$ at 40 K, corresponding to a deviation of $-10$~\% from the experimental value. Following the AF-FM energy difference, the calculation without the Hubbard term results in the smallest T$_C$, underestimating the experimental value by $-32$~\%. In contrast, incorporation of the U parameter for Cr ($U_{Cr} \neq 0$) leads to an overestimation of T$_C$ by $2$~% for $U_{I} = 0$ and by $8$~% for $U_{I} = 2.0$ eV compared to the experimental value.
Magnetic moments with $U_{Cr} = 0$ exhibit the lowest values, yet they follow a similar trend as in previous calculations. The calculated magnetic moments (for Cr, I, and the total magnetic moment of the system) obtained using HSE06 and $U_{Cr} \neq 0$ are similar to previously reported values \cite{kumar2019magnetism, wang2016doping}. 

The band gap for spin-up electrons calculated with the HSE06 functional is the largest, at 1.95 eV, which correlates with previous findings \cite{zhang2018strong,zhang2015robust}. Without the U parameter $U = 0$, the up-spin gap is 1.19 eV. With $U_{Cr} \neq 0$, the up-spin gap is approximately 1.1 eV.
The band gap for spin-down electrons is the largest when calculated using the HSE06 functional, with a value of 3.69 eV, consistent with the findings of earlier reports \cite{zhang2018strong}. This is closely followed by the calculation with $U_{Cr} \neq 0$, which yields a gap of approximately 3.1 eV. When $U = 0$, the gap is 1.9 eV, in agreement with other reports \cite{zhang2018strong,guo2018half,wang2016doping,yang2021enhancing}.
When the energy scaling factor ($\mathcal{E}$) is considered to align the DOS more closely with the HSE06 results, the DFT+U band gaps increase. In particular, the spin-up/spin-down band gaps for $U = 0$ adjust to 1.3/2.1 eV; for $U_{Cr,I} \neq 0$, these band gaps rise to 1.2/3.5 eV.

Our previously discussed results demonstrate the critical role of Hubbard-U repulsion terms in DFT calculations to achieve results similar to those obtained with more demanding methods.
Since the value of the Hubbard-U repulsion depends on the geometry, a possible refinement involves using crystal field theory for symmetry and charge distribution considerations. Therefore, we could apply one repulsion parameter for in-plane orbitals and another for out-of-plane orbitals. However, this sophisticated approach increases the number of free parameters to be tuned, leading to the same technical problems faced by DFT+U+V type approximations \cite{campo2010extended,yu2023active}. Such further refinement is beyond the scope of the present work and is left for future investigations.

\subsubsection{Magnetic Anisotropy Energy}
\begin{table}
\begin{center}
\caption{
Role of spin-orbit coupling on various magnetic properties of CrI$_3$ calculated with different parameters. Band gap, magnetic anisotropy energy (MAE~=~$E_{out-plane}~-~E_{in-plane}$, where MAE $> 0$ indicates an out-of-plane easy axis), and z-projected magnetic moments (in Cr and total). The moment projections in the x and y directions tend to 0.}
\begin{tabular}{ c     c     c     c     c     c     c     c     c     c   } 
\hline 
\hline 
   $U_{Cr_d,I_p}$ & gap &  MAE & m$_{Cr}$  & m$_{T}$  \\
   (eV)& (eV) & (meV/f.u.) & ($\mu_{B}$) &   ($\mu_{B}$)\\ 
\hline 
 \hline 
    \textbf{HSE06} & $1.55$ & $1.26$ & $3.24$ &  $5.80$\\  
    $\left\lbrace 0.0\,,0.0 \right\rbrace$ &  $0.87$ & $1.24$ & $3.02$   &$5.60$\\  
	$\left\lbrace 3.0\,,0.0 \right\rbrace$ &  $0.69$& $1.48$ & $3.40$   &$5.95$\\   
	$\mathbf{\left\lbrace 3.5\,,2.0 \right\rbrace }$ & $0.64$& $1.75$ & $3.45$ &  $5.90$ \\ \hline
\hline 
\end{tabular}
\label{tab:ALL3}
\end{center}
\end{table}

The incorporation of spin-orbit effects modifies the physical and magnetic properties of the CrI$_3$ monolayer, as detailed in Table \ref{tab:ALL3}. These variations can be attributed to a change in the angular momentum of the electrons, which in turn leads to a change in the electronic structure of the atoms by breaking the degeneracy of energy levels. The band gap is reduced when the HSE06+SOC approach is used, decreasing to 1.55 eV (approximately 0.5 eV less than in the absence of SOC). In the case of DFT calculations, the gap reduction is greater, almost half the value obtained without SOC.

The magnetic anisotropy energy (MAE) measures the magnetization preference for a particular direction. Our results indicate a preference for an out-of-plane axis. The HSE06 and DFT calculations with $U_{Cr,I}=\left\lbrace 0.0,\, 0.0 \right\rbrace$ yield comparable values, about 1.2 meV per formula unit, suggesting that the geometry of the system significantly impacts MAE, and may not be a suitable quantity for fitting the U value.
When setting $U_{\text{Cr}}\neq 0$, the MAE increases, reaching 1.5 meV per formula unit for $ U_{Cr,I}=\left\lbrace 3.5,\,0.0 \right\rbrace$. When considering the U in the p-orbitals of I, $ U_{Cr,I}=\left\lbrace 3.5,\, 2.0 \right\rbrace$, the MAE increases to 1.8 meV per formula unit. 
Our findings on magnetic anisotropy are in agreement with previous studies, proposing a MAE of about $1.6$ meV per formula unit. \cite{liu2018analysis,webster2018strain,lado2017origin}. 

 Furthermore, we analyze the response of the monolayer to a biaxial strain for the two sets of U parameters. The energies versus strain are shown in Fig. S5. The results obtained from both models indicate a transition from AFM to FM magnetic ordering when the cell is compressed beyond -6\%.  This finding is in close agreement with previous results, which demonstrated that compressing the cell beyond -6\% results in a change in magnetic ordering \cite{leon2020strain}. 
 
Building upon the robust characterization of the monolayer system, we extend our analysis to CrI$_3$ bilayers, where interlayer interactions and stacking orders play a crucial role in determining the magnetic properties observed experimentally.

\subsubsection{Magnetic ordering and stackings in CrI$_3$ bilayer}

%We have studied the influence of the Hubbard U parameter on the d orbitals of chromium (d-Cr) and the p orbitals of iodine (p-I) and how it modifies the structural, electronic, and magnetic properties of the monolayer CrI$_3$. 
We study now how the Hubbard U terms affect the properties of the bilayer CrI$_3$. 
The CrI$_3$ bilayer exhibits mainly two distinct stacking configurations: the monoclinic structure, which is associated with high temperatures (HT, space group $C2/m$), and the rhombohedral structure, which is observed at low temperatures (LT, space group $R\Bar{3}$) \cite{morell2019control,leon2020strain}. 

In the monolayer, the two Cr atoms in the unit cell can be coupled ferromagnetically ($\uparrow\uparrow$) or antiferromagnetically ($\uparrow\downarrow$). However, in the bilayer, the coupling landscape of magnetic moments is more complex. The mentioned complexity gives rise to discrepancies in the results, which depend on the calculation methodology used.
Experimental evidence indicates that the HT structure displays ferromagnetic behavior within the layers and antiferromagnetic between the layers. 
%with the spins aligned either parallel or antiparallel to each other
($\uparrow\uparrow/\downarrow\downarrow$ or $\downarrow\downarrow/\uparrow\uparrow$). 
In contrast, the LT structure exhibits ferromagnetic behavior within and between layers ($\uparrow\uparrow/\uparrow\uparrow$ or $\downarrow\downarrow/\downarrow\downarrow)$ \cite{huang2018electrical,mcguire2015coupling}.

\begin{figure}
\includegraphics[clip,width=0.95\columnwidth,clip]{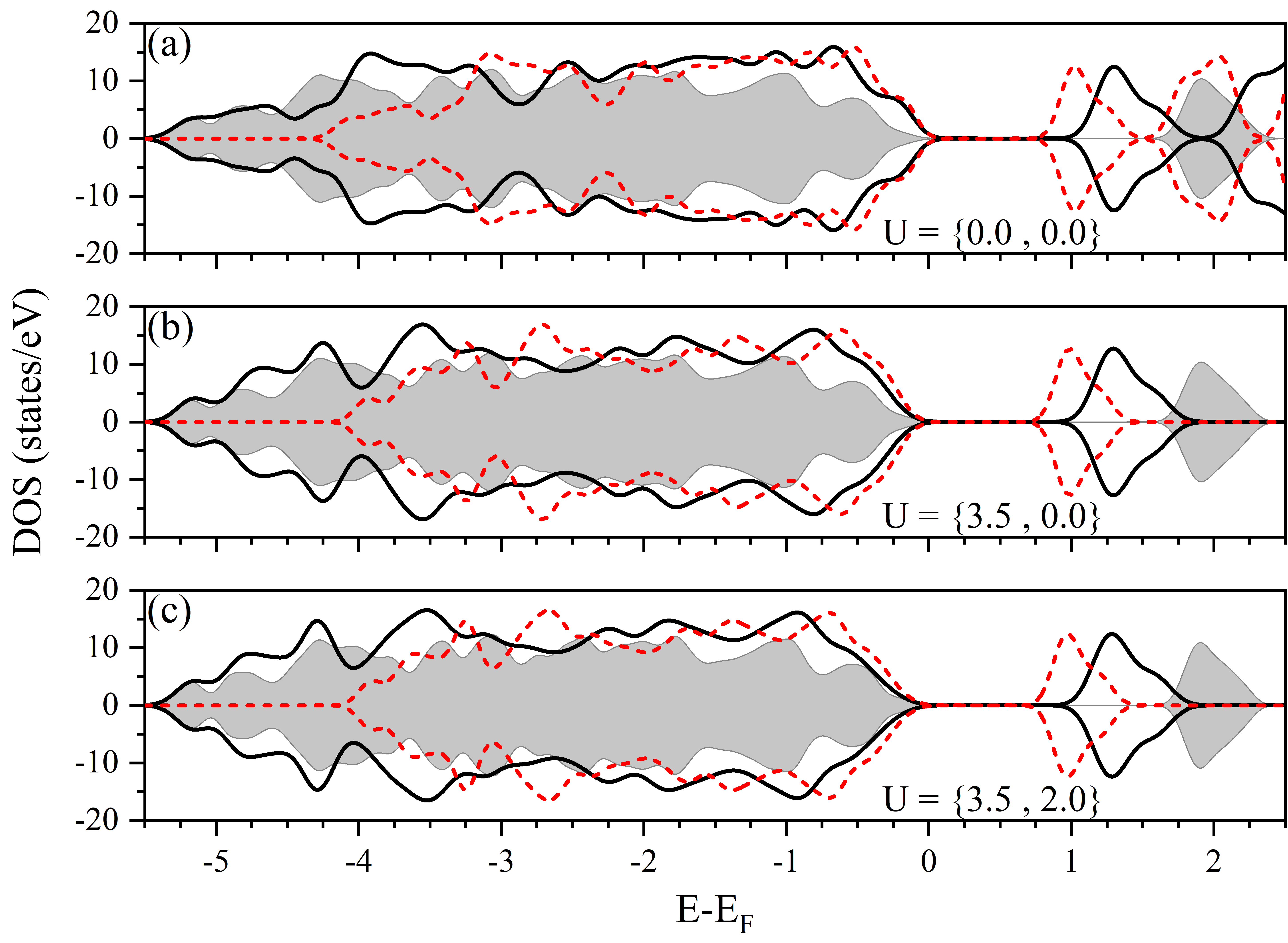}
    \caption{Density of states of CrI$_3$ bilayer in the HT-AF stacking within different models. The hybrid HSE06 functional is our reference, shown as a shadow region. 
    The raw DFT+U density of states is represented by a dashed red line, and the energy-scaled DFT+U is represented by a black line. 
    In panel (a), results for $U_{Cr,I}=\left\lbrace 0.0\,,0.0 \right\rbrace$, in (b) $U_{Cr,I}=\left\lbrace 3.5\,,0.0 \right\rbrace$, and (c) $U_{Cr,I}=\left\lbrace 3.5\,,2.0 \right\rbrace$.}
    \label{fig:DOSCrI3-Bilayer}
\end{figure}

%%% Esquema bicapa, creo que no es necesaria
%\begin{figure}
%\includegraphics[clip,width=0.85\columnwidth,clip]{Figures/fig7.png}
%    \caption{
%    Caption
%    }
%    \label{fig:Bicapa-Top-Side}
%\end{figure}

\begin{table*}[]
\centering
\caption{Bilayer CrI$_3$ in HT and LT stacking. Difference between  AF and FM interlayer magnetic coupling. Lattice constant, Cr-Cr interlayer distance, and band gap (gap $\uparrow, \downarrow$). The energy differences in red are in disagreement with the experimental reports.} \label{tab:bilayer}
\begin{tabular}{c    c    c    c    c    c    c    c }
\hline \hline
Configuration & Stacking & \begin{tabular}[c]{@{}c@{}}$\uparrow\uparrow/\downarrow\downarrow - \uparrow\uparrow/\uparrow\uparrow$\\ (meV/Cr)\end{tabular} & Magnetism & \begin{tabular}[c]{@{}c@{}}    a     \\ (\AA{})\end{tabular} & \begin{tabular}[c]{@{}c@{}}d\\ (\AA{})\end{tabular} & \begin{tabular}[c]{@{}c@{}}gap $\uparrow$\\ (eV)\end{tabular} & \begin{tabular}[c]{@{}c@{}}gap $\downarrow$\\ (eV)\end{tabular} \\ \hline \hline
\multirow{4}{*}{\textbf{HSE06}} & \multirow{2}{*}{HT} & \multirow{2}{*}{$-0.335$} & $\uparrow\uparrow/\uparrow\uparrow$ & 6.79 & 6.47 & 1.77 & 3.90 \\ \cline{4-8} 
 &  &  & $\uparrow\uparrow/\downarrow\downarrow$ & 6.79 & 6.47 & 1.87 & 1.87 \\ \cline{2-8} 
 & \multirow{2}{*}{LT} & \multirow{2}{*}{3.502} & $\uparrow\uparrow/\uparrow\uparrow$ & 6.80 & 6.42 & 1.80 & 3.44 \\ \cline{4-8} 
 &  &  & $\uparrow\uparrow/\downarrow\downarrow$ & 6.79 & 6.43 & 1.92 & 1.92 \\ \hline 
\multirow{4}{*}{$\left\lbrace 0.0\,,0.0 \right\rbrace$} & \multirow{2}{*}{HT} & \multirow{2}{*}{$\color{red}{0.286}\color{black}$} & $\uparrow\uparrow/\uparrow\uparrow$ & 7.11 & 6.82 & 0.98 & 2.04 \\ \cline{4-8} 
 &  &  & $\uparrow\uparrow/\downarrow\downarrow$ & 7.11 & 6.83 & 1.09 & 1.09 \\ \cline{2-8} 
 & \multirow{2}{*}{LT} & \multirow{2}{*}{1.183} & $\uparrow\uparrow/\uparrow\uparrow$ & 7.20 & 6.72 & 1.03 & 2.03 \\ \cline{4-8} 
 &  &  & $\uparrow\uparrow/\downarrow\downarrow$ & 7.18 & 6.72 & 1.09 & 1.09 \\ \hline 
\multirow{4}{*}{$\left\lbrace 3.5\,,0.0 \right\rbrace$} & \multirow{2}{*}{HT} & \multirow{2}{*}{-0.484} & $\uparrow\uparrow/\uparrow\uparrow$ & 7.18 & 7.03 & 0.93 & 3.24 \\ \cline{4-8} 
 &  &  & $\uparrow\uparrow/\downarrow\downarrow$ & 7.20 & 7.00 & 1.04 & 1.04 \\ \cline{2-8} 
 & \multirow{2}{*}{LT} & \multirow{2}{*}{2.464} & $\uparrow\uparrow/\uparrow\uparrow$ & 7.10 & 7.04 & 0.99 & 3.29 \\ \cline{4-8} 
 &  &  & $\uparrow\uparrow/\downarrow\downarrow$ & 7.10 & 6.93 & 1.04 & 1.04 \\ \hline 
\multirow{4}{*}{$\mathbf{\left\lbrace 3.5\,,2.0 \right\rbrace }$ } & \multirow{2}{*}{HT} & \multirow{2}{*}{-0.597} & $\uparrow\uparrow/\uparrow\uparrow$ & 7.19 & 7.03 & 0.91 & 3.34 \\ \cline{4-8} 
 &  &  & $\uparrow\uparrow/\downarrow\downarrow$ & 7.21 & 6.99 & 1.03 & 1.03 \\ \cline{2-8} 
 & \multirow{2}{*}{LT} & \multirow{2}{*}{1.107} & $\uparrow\uparrow/\uparrow\uparrow$ & 7.20 & 6.94 & 0.97 & 3.38 \\ \cline{4-8} 
 &  &  & $\uparrow\uparrow/\downarrow\downarrow$ & 7.20 & 6.98 & 1.02 & 1.02 \\ \hline \hline
\end{tabular}
\end{table*}

Including non-local van der Waals (vdW) corrections becomes essential to correctly describe magnetism in HT stacking due to the critical role of weak but long-ranged super-superexchange magnetic interactions in determining the antiferromagnetic character of interlayer interactions.

We conducted several benchmarks to evaluate the most accurate computational method for capturing these interactions and selected the vdW-DF approach \cite{thonhauser2007van,dion2004van,roman2009efficient,klimevs2011van}. In contrast, previous calculations that employed local vdW corrections using the DFT + D2, DFT + D3 (zero damping) and DFT + D3 (BJ) methods led to results that were inconsistent with experimental observations \cite{dion2004van,roman2009efficient}.
The nonlocal van der Waals (vdW-DF) correction is currently incompatible with hybrid functionals, and therefore calculations that include both approaches are today unfeasible\cite{Nonlocal}.
The hybrid results presented are obtained using the hybrid HSE06 functional and are based on single point calculations.  These calculations are then performed within the van der Waals correction vdW-D3 (BJ) using the relaxed geometry calculated with DFT+U for $U_{Cr,I}=\left\lbrace0.0, 0.0 \right\rbrace$. 
Table \ref{tab:bilayer} summarizes the results for the bilayer in both HT and LT stackings.

The lattice constant for LT and HT stacking configurations increases with the Hubbard value due to differences in the in-plane van der Waals corrections \cite{klimevs2011van}, the lattice constant obtained is approximately 4 \% larger compared to previous theoretical reports; however, the distance between layers remains similar to previous reports \cite{jiang2019stacking,leon2020strain,guo2020enhanced}.

It is important to note that despite using vdw-DF, the setting U=$\left\lbrace 0.0,0.0 \right\rbrace$ cannot predict the AF character of the interaction between the HT layers.
In contrast, the settings with $U_{Cr,I}=\left\lbrace3.5, 0.0 \right\rbrace$ and $U_{Cr,I}=\left\lbrace3.5, 2.0 \right\rbrace$ accurately capture the experimentally observed out-of-plane antiferromagnetism. As the interlayer distance, d, remains constant, introducing nonzero Hubbard U corrections for chromium and iodine improves the antiferromagnetic nature of the interlayer coupling. Specifically, the AF character in the $U_{Cr,I}=\left\lbrace3.5, 2.0 \right\rbrace$ configuration is approximately 1.4 times higher than in the $U_{Cr,I}=\left\lbrace3.5, 0.0 \right\rbrace$ configuration. 
In the LT stacking, the energy differences between the AF and FM states for $U_{Cr,I}=\left\lbrace 0.0,\,0.0 \right\rbrace$ and $U_{Cr,I}=\left\lbrace3.5, 2.0 \right\rbrace$ are comparable. However, the energy difference for $U_{Cr,I}=\left\lbrace 3.5,\,0.0 \right\rbrace$ increases by a factor of two.  
The use of a U parameter in the iodine p-orbitals tends to favor AF interlayer interactions, offering a more accurate picture of the stacking-dependent magnetic behavior in CrI$_3$ bilayers\cite{jiang2019stacking,leon2020strain,morell2019control,guo2020enhanced}. This improvement is achieved by using the two U parameters on Cr and I and in conjunction with the most recent van der Waals functional.

%%%% NEW
Figure \ref{fig:DOSCrI3-Bilayer} shows the density of states (DOS) of the CrI$3$ bilayer calculated using different Hubbard-U parameter configurations, compared against the reference hybrid functional HSE06 results (shaded area). The dashed red line represents the raw DFT+U density of states, while the solid black line indicates the energy-scaled DFT+U. We consider three different configurations: Fig. \ref{fig:DOSCrI3-Bilayer}(a) without Hubbard corrections ($U_{Cr,I} = \left\lbrace 0.0, 0.0 \right\rbrace$), Fig. \ref{fig:DOSCrI3-Bilayer}(b) with corrections only for the Cr 3d orbitals ($U_{Cr,I} = \left\lbrace 3.5, 0.0 \right\rbrace$), and Fig. \ref{fig:DOSCrI3-Bilayer}(c) incorporating Hubbard corrections for both Cr 3d and I 5p orbitals ($U_{Cr,I} = \left\lbrace 3.5, 2.0 \right\rbrace$).

Comparing Figs. \ref{fig:DOSCrI3-Bilayer}(b) and (c), we notice that the addition of Hubbard corrections to iodine orbitals substantially improves alignment with the hybrid functional HSE06 DOS. Among all configurations tested, the best agreement with the HSE06 reference (including vdW-D3(BJ) corrections) is achieved using Hubbard corrections for the chromium and iodine orbitals, specifically $U_{Cr, I}=\left\lbrace 3.5, 2.0 \right\rbrace$, after applying the energy scaling.

It is important to highlight that current technical limitations prevent combining non-local van der Waals corrections (vdW-DF) with hybrid functional approaches in the same calculation, making it impractical to include both corrections simultaneously. Thus, the proposed DFT+U approach with carefully calibrated Hubbard parameters presents a highly advantageous alternative for accurately describing the electronic structure of CrI$_3$ bilayers at a considerably lower computational cost.

Previous DFT studies employing Grimme and Becke-Johnson van der Waals corrections reported band gaps between 0.5 and 0.6 eV for HT and LT stackings. However, the gaps identified in our DFT analysis are closer to, although slightly lower than, those obtained from HSE06 calculations \cite{jiang2019stacking,leon2020strain}. As expected, the energy band gap for both spin components is identical in nanostructures exhibiting AF coupling. In the case of the HT stacking, the HSE06 calculation shows a gap of 1.9 eV. In contrast, our DFT+U calculations yield approximately 1.1 eV (which increases to 1.3 eV after applying the energy scaling factor $\varepsilon$), with minor variations between the tested sets of DFT + U parameters.

Moreover, Fig. S6 in the supplemental material explicitly demonstrates how the magnetic interactions between layers respond to changes in interlayer distance, providing important insights into the stability of magnetic ordering under strain. Starting from the equilibrium interlayer distance, we calculate the energy difference between antiferromagnetic ($\uparrow \uparrow / \downarrow \downarrow$) and ferromagnetic ($\uparrow \uparrow / \uparrow \uparrow$) couplings. For inward displacements below equilibrium, antiferromagnetic coupling remains preferred. In contrast, for outward displacements beyond 0.2 \AA{} from equilibrium, the configuration with $U_{Cr, I}=\left\lbrace 3.5, 2.0 \right\rbrace$ favors ferromagnetic coupling. This behavior aligns well with the earlier discussions in this paper, stressing the sensitivity of magnetic properties to Hubbard parameter tuning.

\section{Final Remarks}

In this work, we have shown that including Hubbard $U$ corrections for chromium 3$d$ and iodine 5$p$ orbitals enhances the accuracy of DFT+U calculations for CrI$_3$ monolayers and bilayers. Through systematic benchmarking against hybrid functional (HSE06) results, we identified optimal Hubbard parameters of $U_{\text{Cr}} = 3.5\,\text{eV}$ and $U_{\text{I}} = 2.0\,\text{eV}$, which show excellent agreement with reference calculations, quantified by a Pearson correlation coefficient $\mathcal{P}=0.95$.

Including Hubbard corrections on iodine orbitals is particularly beneficial for accurately capturing the delicate super-exchange and super-super-exchange interactions responsible for interlayer magnetic coupling, which is essential for interpreting experimental observations. This improved theoretical description represents a meaningful advancement towards reliably modeling CrI$_3$ bilayers, especially for the experimentally relevant high-temperature (HT) stacking configuration.

The computational approach presented here facilitates the accurate study of CrI$_3$ and establishes a practical methodological framework that is broadly applicable to other magnetic-layered systems exhibiting strong electronic correlations. Our results provide a clear and transferable strategy for calibrating Hubbard parameters, thereby improving theoretical predictions of electronic and magnetic properties in diverse correlated materials.\\

\section*{Declaration of competing interest}
The authors declare that they have no known competing financial interests or personal relationships that could have appeared to influence the work reported in this paper.

\section*{Data availability}
All data supporting the conclusions of this study are provided in the supplementary information files. Additional datasets are available from the corresponding author upon reasonable request.

\section*{{CRediT} authorship contribution statement}
\textbf{D. Lauer}: Investigation, Visualization, and Writing.
\textbf{J. W. Gonz\'{a}lez}: Methodology, Conceptualization, Validation, and Writing.
\textbf{E. Su\'{a}rez Morell}: Methodology, Conceptualization, Validation, and Writing.
\textbf{A. Ayuela}: Methodology, Conceptualization, Validation, and Writing.
All authors provided critical feedback, contributed to the design, analysis, and writing of the investigation, and approved the final version of the manuscript.

\section*{acknowledgment}
DL and JWG would like to acknowledge DIPC's hospitality during the completion of this work.
JWG acknowledges financial support from ANID-FONDECY grant N. 1220700 (Chile). 
ESM and JWG acknowledge the financial support from ANID-FONDECYT 1221301 (Chile).
AA and DL acknowledge funding from the Spanish Ministry of Science and Innovation (grants nos. PID2019-105488GB-I00,  PID2022-139230NB-I00, and TED2021-132074B-C32), the Gobierno Vasco UPV/EHU (project no. IT-1569-22), the Diputación Foral de Gipuzkoa (Project No. 2023-CIEN-000077-01), MIRACLE project (GA 964450), and NaturSea-PV (GA 101084348). Research conducted in the scope of the Transnational Common Laboratory (LTC) Aquitaine-Euskadi Network in Green Concrete and Cement-based Materials.

%\bibliography{file_bib}% Produces the bibliography via BibTeX.

\begin{thebibliography}{51}%
\makeatletter
\providecommand \@ifxundefined [1]{%
 \@ifx{#1\undefined}
}%
\providecommand \@ifnum [1]{%
 \ifnum #1\expandafter \@firstoftwo
 \else \expandafter \@secondoftwo
 \fi
}%
\providecommand \@ifx [1]{%
 \ifx #1\expandafter \@firstoftwo
 \else \expandafter \@secondoftwo
 \fi
}%
\providecommand \natexlab [1]{#1}%
\providecommand \enquote  [1]{``#1''}%
\providecommand \bibnamefont  [1]{#1}%
\providecommand \bibfnamefont [1]{#1}%
\providecommand \citenamefont [1]{#1}%
\providecommand \href@noop [0]{\@secondoftwo}%
\providecommand \href [0]{\begingroup \@sanitize@url \@href}%
\providecommand \@href[1]{\@@startlink{#1}\@@href}%
\providecommand \@@href[1]{\endgroup#1\@@endlink}%
\providecommand \@sanitize@url [0]{\catcode `\\12\catcode `\$12\catcode
  `\&12\catcode `\#12\catcode `\^12\catcode `\_12\catcode `\%12\relax}%
\providecommand \@@startlink[1]{}%
\providecommand \@@endlink[0]{}%
\providecommand \url  [0]{\begingroup\@sanitize@url \@url }%
\providecommand \@url [1]{\endgroup\@href {#1}{\urlprefix }}%
\providecommand \urlprefix  [0]{URL }%
\providecommand \Eprint [0]{\href }%
\providecommand \doibase [0]{https://doi.org/}%
\providecommand \selectlanguage [0]{\@gobble}%
\providecommand \bibinfo  [0]{\@secondoftwo}%
\providecommand \bibfield  [0]{\@secondoftwo}%
\providecommand \translation [1]{[#1]}%
\providecommand \BibitemOpen [0]{}%
\providecommand \bibitemStop [0]{}%
\providecommand \bibitemNoStop [0]{.\EOS\space}%
\providecommand \EOS [0]{\spacefactor3000\relax}%
\providecommand \BibitemShut  [1]{\csname bibitem#1\endcsname}%
\let\auto@bib@innerbib\@empty
%</preamble>
\bibitem [{\citenamefont {Huang}\ \emph {et~al.}(2017)\citenamefont {Huang},
  \citenamefont {Clark}, \citenamefont {Navarro-Moratalla}, \citenamefont
  {Klein}, \citenamefont {Cheng}, \citenamefont {Seyler}, \citenamefont
  {Zhong}, \citenamefont {Schmidgall}, \citenamefont {McGuire}, \citenamefont
  {Cobden} \emph {et~al.}}]{huang2017layer}%
  \BibitemOpen
  \bibfield  {author} {\bibinfo {author} {\bibfnamefont {B.}~\bibnamefont
  {Huang}}, \bibinfo {author} {\bibfnamefont {G.}~\bibnamefont {Clark}},
  \bibinfo {author} {\bibfnamefont {E.}~\bibnamefont {Navarro-Moratalla}},
  \bibinfo {author} {\bibfnamefont {D.~R.}\ \bibnamefont {Klein}}, \bibinfo
  {author} {\bibfnamefont {R.}~\bibnamefont {Cheng}}, \bibinfo {author}
  {\bibfnamefont {K.~L.}\ \bibnamefont {Seyler}}, \bibinfo {author}
  {\bibfnamefont {D.}~\bibnamefont {Zhong}}, \bibinfo {author} {\bibfnamefont
  {E.}~\bibnamefont {Schmidgall}}, \bibinfo {author} {\bibfnamefont {M.~A.}\
  \bibnamefont {McGuire}}, \bibinfo {author} {\bibfnamefont {D.~H.}\
  \bibnamefont {Cobden}}, \emph {et~al.},\ }\bibfield  {title} {\bibinfo
  {title} {Layer-dependent ferromagnetism in a {Van} der {Waals} crystal down
  to the monolayer limit},\ }\href@noop {} {\bibfield  {journal} {\bibinfo
  {journal} {Nature}\ }\textbf {\bibinfo {volume} {546}},\ \bibinfo {pages}
  {270} (\bibinfo {year} {2017})}\BibitemShut {NoStop}%
\bibitem [{\citenamefont {Liu}\ \emph {et~al.}(2016)\citenamefont {Liu},
  \citenamefont {Sun}, \citenamefont {Kawazoe},\ and\ \citenamefont
  {Jena}}]{liu2016exfoliating}%
  \BibitemOpen
  \bibfield  {author} {\bibinfo {author} {\bibfnamefont {J.}~\bibnamefont
  {Liu}}, \bibinfo {author} {\bibfnamefont {Q.}~\bibnamefont {Sun}}, \bibinfo
  {author} {\bibfnamefont {Y.}~\bibnamefont {Kawazoe}},\ and\ \bibinfo {author}
  {\bibfnamefont {P.}~\bibnamefont {Jena}},\ }\bibfield  {title} {\bibinfo
  {title} {Exfoliating biocompatible ferromagnetic {Cr-trihalide} monolayers},\
  }\href@noop {} {\bibfield  {journal} {\bibinfo  {journal} {Physical Chemistry
  Chemical Physics}\ }\textbf {\bibinfo {volume} {18}},\ \bibinfo {pages}
  {8777} (\bibinfo {year} {2016})}\BibitemShut {NoStop}%
\bibitem [{\citenamefont {Yang}\ \emph {et~al.}(2021)\citenamefont {Yang},
  \citenamefont {Hu}, \citenamefont {Shen}, \citenamefont {Krasheninnikov},
  \citenamefont {Chen},\ and\ \citenamefont {Sun}}]{yang2021enhancing}%
  \BibitemOpen
  \bibfield  {author} {\bibinfo {author} {\bibfnamefont {Q.}~\bibnamefont
  {Yang}}, \bibinfo {author} {\bibfnamefont {X.}~\bibnamefont {Hu}}, \bibinfo
  {author} {\bibfnamefont {X.}~\bibnamefont {Shen}}, \bibinfo {author}
  {\bibfnamefont {A.~V.}\ \bibnamefont {Krasheninnikov}}, \bibinfo {author}
  {\bibfnamefont {Z.}~\bibnamefont {Chen}},\ and\ \bibinfo {author}
  {\bibfnamefont {L.}~\bibnamefont {Sun}},\ }\bibfield  {title} {\bibinfo
  {title} {Enhancing {Ferromagnetism} and tuning electronic properties of
  {CrI$_3$} monolayers by adsorption of transition-metal atoms},\ }\href@noop
  {} {\bibfield  {journal} {\bibinfo  {journal} {ACS Applied Materials \&
  Interfaces}\ }\textbf {\bibinfo {volume} {13}},\ \bibinfo {pages} {21593}
  (\bibinfo {year} {2021})}\BibitemShut {NoStop}%
\bibitem [{\citenamefont {Wang}\ \emph {et~al.}(2011)\citenamefont {Wang},
  \citenamefont {Eyert},\ and\ \citenamefont
  {Schwingenschl{\"o}gl}}]{wang2011electronic}%
  \BibitemOpen
  \bibfield  {author} {\bibinfo {author} {\bibfnamefont {H.}~\bibnamefont
  {Wang}}, \bibinfo {author} {\bibfnamefont {V.}~\bibnamefont {Eyert}},\ and\
  \bibinfo {author} {\bibfnamefont {U.}~\bibnamefont {Schwingenschl{\"o}gl}},\
  }\bibfield  {title} {\bibinfo {title} {Electronic structure and magnetic
  ordering of the semiconducting chromium trihalides {CrCl$_3$}, {CrBr$_3$},
  and {CrI$_3$}},\ }\href@noop {} {\bibfield  {journal} {\bibinfo  {journal}
  {Journal of Physics: Condensed Matter}\ }\textbf {\bibinfo {volume} {23}},\
  \bibinfo {pages} {116003} (\bibinfo {year} {2011})}\BibitemShut {NoStop}%
\bibitem [{\citenamefont {Zhang}\ \emph {et~al.}(2015)\citenamefont {Zhang},
  \citenamefont {Qu}, \citenamefont {Zhu},\ and\ \citenamefont
  {Lam}}]{zhang2015robust}%
  \BibitemOpen
  \bibfield  {author} {\bibinfo {author} {\bibfnamefont {W.-B.}\ \bibnamefont
  {Zhang}}, \bibinfo {author} {\bibfnamefont {Q.}~\bibnamefont {Qu}}, \bibinfo
  {author} {\bibfnamefont {P.}~\bibnamefont {Zhu}},\ and\ \bibinfo {author}
  {\bibfnamefont {C.-H.}\ \bibnamefont {Lam}},\ }\bibfield  {title} {\bibinfo
  {title} {Robust intrinsic ferromagnetism and half semiconductivity in stable
  two-dimensional single-layer chromium trihalides},\ }\href@noop {} {\bibfield
   {journal} {\bibinfo  {journal} {Journal of Materials Chemistry C}\ }\textbf
  {\bibinfo {volume} {3}},\ \bibinfo {pages} {12457} (\bibinfo {year}
  {2015})}\BibitemShut {NoStop}%
\bibitem [{\citenamefont {McGuire}\ \emph {et~al.}(2015)\citenamefont
  {McGuire}, \citenamefont {Dixit}, \citenamefont {Cooper},\ and\ \citenamefont
  {Sales}}]{mcguire2015coupling}%
  \BibitemOpen
  \bibfield  {author} {\bibinfo {author} {\bibfnamefont {M.~A.}\ \bibnamefont
  {McGuire}}, \bibinfo {author} {\bibfnamefont {H.}~\bibnamefont {Dixit}},
  \bibinfo {author} {\bibfnamefont {V.~R.}\ \bibnamefont {Cooper}},\ and\
  \bibinfo {author} {\bibfnamefont {B.~C.}\ \bibnamefont {Sales}},\ }\bibfield
  {title} {\bibinfo {title} {Coupling of crystal structure and magnetism in the
  layered, ferromagnetic insulator {CrI$_3$}},\ }\href@noop {} {\bibfield
  {journal} {\bibinfo  {journal} {Chemistry of Materials}\ }\textbf {\bibinfo
  {volume} {27}},\ \bibinfo {pages} {612} (\bibinfo {year} {2015})}\BibitemShut
  {NoStop}%
\bibitem [{\citenamefont {Lado}\ and\ \citenamefont
  {Fern{\'a}ndez-Rossier}(2017)}]{lado2017origin}%
  \BibitemOpen
  \bibfield  {author} {\bibinfo {author} {\bibfnamefont {J.~L.}\ \bibnamefont
  {Lado}}\ and\ \bibinfo {author} {\bibfnamefont {J.}~\bibnamefont
  {Fern{\'a}ndez-Rossier}},\ }\bibfield  {title} {\bibinfo {title} {On the
  origin of magnetic anisotropy in two dimensional {CrI$_3$}},\ }\href@noop {}
  {\bibfield  {journal} {\bibinfo  {journal} {2D Materials}\ }\textbf {\bibinfo
  {volume} {4}},\ \bibinfo {pages} {035002} (\bibinfo {year}
  {2017})}\BibitemShut {NoStop}%
\bibitem [{\citenamefont {Garza}\ and\ \citenamefont
  {Scuseria}(2016)}]{garza2016predicting}%
  \BibitemOpen
  \bibfield  {author} {\bibinfo {author} {\bibfnamefont {A.~J.}\ \bibnamefont
  {Garza}}\ and\ \bibinfo {author} {\bibfnamefont {G.~E.}\ \bibnamefont
  {Scuseria}},\ }\bibfield  {title} {\bibinfo {title} {Predicting band gaps
  with hybrid density functionals},\ }\href@noop {} {\bibfield  {journal}
  {\bibinfo  {journal} {The journal of physical chemistry letters}\ }\textbf
  {\bibinfo {volume} {7}},\ \bibinfo {pages} {4165} (\bibinfo {year}
  {2016})}\BibitemShut {NoStop}%
\bibitem [{\citenamefont {Dudarev}\ \emph {et~al.}(1998)\citenamefont
  {Dudarev}, \citenamefont {Botton}, \citenamefont {Savrasov}, \citenamefont
  {Humphreys},\ and\ \citenamefont {Sutton}}]{dudarev1998electron}%
  \BibitemOpen
  \bibfield  {author} {\bibinfo {author} {\bibfnamefont {S.~L.}\ \bibnamefont
  {Dudarev}}, \bibinfo {author} {\bibfnamefont {G.~A.}\ \bibnamefont {Botton}},
  \bibinfo {author} {\bibfnamefont {S.~Y.}\ \bibnamefont {Savrasov}}, \bibinfo
  {author} {\bibfnamefont {C.}~\bibnamefont {Humphreys}},\ and\ \bibinfo
  {author} {\bibfnamefont {A.~P.}\ \bibnamefont {Sutton}},\ }\bibfield  {title}
  {\bibinfo {title} {Electron-energy-loss spectra and the structural stability
  of nickel oxide: {An} {LSDA+ U} study},\ }\href@noop {} {\bibfield  {journal}
  {\bibinfo  {journal} {Physical Review B}\ }\textbf {\bibinfo {volume} {57}},\
  \bibinfo {pages} {1505} (\bibinfo {year} {1998})}\BibitemShut {NoStop}%
\bibitem [{\citenamefont {Krukau}\ \emph {et~al.}(2006)\citenamefont {Krukau},
  \citenamefont {Vydrov}, \citenamefont {Izmaylov},\ and\ \citenamefont
  {Scuseria}}]{krukau2006influence}%
  \BibitemOpen
  \bibfield  {author} {\bibinfo {author} {\bibfnamefont {A.~V.}\ \bibnamefont
  {Krukau}}, \bibinfo {author} {\bibfnamefont {O.~A.}\ \bibnamefont {Vydrov}},
  \bibinfo {author} {\bibfnamefont {A.~F.}\ \bibnamefont {Izmaylov}},\ and\
  \bibinfo {author} {\bibfnamefont {G.~E.}\ \bibnamefont {Scuseria}},\
  }\bibfield  {title} {\bibinfo {title} {Influence of the exchange screening
  parameter on the performance of screened hybrid functionals},\ }\href@noop {}
  {\bibfield  {journal} {\bibinfo  {journal} {The Journal of Chemical Physics}\
  }\textbf {\bibinfo {volume} {125}} (\bibinfo {year} {2006})}\BibitemShut
  {NoStop}%
\bibitem [{\citenamefont {Ibragimova}\ \emph {et~al.}(2018)\citenamefont
  {Ibragimova}, \citenamefont {Ganchenkova}, \citenamefont {Karazhanov},\ and\
  \citenamefont {Marstein}}]{ibragimova2018first}%
  \BibitemOpen
  \bibfield  {author} {\bibinfo {author} {\bibfnamefont {R.}~\bibnamefont
  {Ibragimova}}, \bibinfo {author} {\bibfnamefont {M.}~\bibnamefont
  {Ganchenkova}}, \bibinfo {author} {\bibfnamefont {S.}~\bibnamefont
  {Karazhanov}},\ and\ \bibinfo {author} {\bibfnamefont {E.~S.}\ \bibnamefont
  {Marstein}},\ }\bibfield  {title} {\bibinfo {title} {First-principles study
  of {SnS} electronic properties using {LDA}, {PBE} and {HSE06} functionals},\
  }\href@noop {} {\bibfield  {journal} {\bibinfo  {journal} {Philosophical
  Magazine}\ }\textbf {\bibinfo {volume} {98}},\ \bibinfo {pages} {710}
  (\bibinfo {year} {2018})}\BibitemShut {NoStop}%
\bibitem [{\citenamefont {Yang}\ and\ \citenamefont
  {Fan}(2017)}]{yang2017understanding}%
  \BibitemOpen
  \bibfield  {author} {\bibinfo {author} {\bibfnamefont {J.}~\bibnamefont
  {Yang}}\ and\ \bibinfo {author} {\bibfnamefont {Q.}~\bibnamefont {Fan}},\
  }\bibfield  {title} {\bibinfo {title} {Understanding electronic and optical
  properties of strontium titanate at both {PBE} and {HSE06} levels},\ }in\
  \href@noop {} {\emph {\bibinfo {booktitle} {IOP Conference Series: Materials
  Science and Engineering}}},\ Vol.\ \bibinfo {volume} {167}\ (\bibinfo
  {organization} {IOP Publishing},\ \bibinfo {year} {2017})\ p.\ \bibinfo
  {pages} {012010}\BibitemShut {NoStop}%
\bibitem [{\citenamefont {Schäfer}\ \emph {et~al.}(2021)\citenamefont
  {Schäfer}, \citenamefont {Daelman},\ and\ \citenamefont
  {Lopez}}]{schafer2021cerium}%
  \BibitemOpen
  \bibfield  {author} {\bibinfo {author} {\bibfnamefont {T.}~\bibnamefont
  {Schäfer}}, \bibinfo {author} {\bibfnamefont {N.}~\bibnamefont {Daelman}},\
  and\ \bibinfo {author} {\bibfnamefont {N.}~\bibnamefont {Lopez}},\ }\bibfield
   {title} {\bibinfo {title} {Cerium oxides without {U}: The role of
  many-electron correlation},\ }\href@noop {} {\bibfield  {journal} {\bibinfo
  {journal} {The Journal of Physical Chemistry Letters}\ }\textbf {\bibinfo
  {volume} {12}},\ \bibinfo {pages} {6277} (\bibinfo {year}
  {2021})}\BibitemShut {NoStop}%
\bibitem [{\citenamefont {Li}\ \emph {et~al.}(2020)\citenamefont {Li},
  \citenamefont {Li}, \citenamefont {B{\"a}umer},\ and\ \citenamefont
  {Moskaleva}}]{li2020assessment}%
  \BibitemOpen
  \bibfield  {author} {\bibinfo {author} {\bibfnamefont {S.}~\bibnamefont
  {Li}}, \bibinfo {author} {\bibfnamefont {Y.}~\bibnamefont {Li}}, \bibinfo
  {author} {\bibfnamefont {M.}~\bibnamefont {B{\"a}umer}},\ and\ \bibinfo
  {author} {\bibfnamefont {L.~V.}\ \bibnamefont {Moskaleva}},\ }\bibfield
  {title} {\bibinfo {title} {Assessment of {PBE+ U} and {HSE06} methods and
  determination of optimal parameter{ U} for the structural and energetic
  properties of rare earth oxides},\ }\href@noop {} {\bibfield  {journal}
  {\bibinfo  {journal} {The Journal of Chemical Physics}\ }\textbf {\bibinfo
  {volume} {153}} (\bibinfo {year} {2020})}\BibitemShut {NoStop}%
\bibitem [{\citenamefont {Pandey}\ \emph {et~al.}(2017)\citenamefont {Pandey},
  \citenamefont {Xu}, \citenamefont {Williamson}, \citenamefont {Nelson},\ and\
  \citenamefont {Li}}]{pandey2017electronic}%
  \BibitemOpen
  \bibfield  {author} {\bibinfo {author} {\bibfnamefont {S.~C.}\ \bibnamefont
  {Pandey}}, \bibinfo {author} {\bibfnamefont {X.}~\bibnamefont {Xu}}, \bibinfo
  {author} {\bibfnamefont {I.}~\bibnamefont {Williamson}}, \bibinfo {author}
  {\bibfnamefont {E.~B.}\ \bibnamefont {Nelson}},\ and\ \bibinfo {author}
  {\bibfnamefont {L.}~\bibnamefont {Li}},\ }\bibfield  {title} {\bibinfo
  {title} {Electronic and vibrational properties of transition metal-oxides:
  {Comparison} of {GGA}, {GGA+ U}, and hybrid approaches},\ }\href@noop {}
  {\bibfield  {journal} {\bibinfo  {journal} {Chemical Physics Letters}\
  }\textbf {\bibinfo {volume} {669}},\ \bibinfo {pages} {1} (\bibinfo {year}
  {2017})}\BibitemShut {NoStop}%
\bibitem [{\citenamefont {Dion}\ \emph {et~al.}(2004)\citenamefont {Dion},
  \citenamefont {Rydberg}, \citenamefont {Schr{\"o}der}, \citenamefont
  {Langreth},\ and\ \citenamefont {Lundqvist}}]{dion2004van}%
  \BibitemOpen
  \bibfield  {author} {\bibinfo {author} {\bibfnamefont {M.}~\bibnamefont
  {Dion}}, \bibinfo {author} {\bibfnamefont {H.}~\bibnamefont {Rydberg}},
  \bibinfo {author} {\bibfnamefont {E.}~\bibnamefont {Schr{\"o}der}}, \bibinfo
  {author} {\bibfnamefont {D.~C.}\ \bibnamefont {Langreth}},\ and\ \bibinfo
  {author} {\bibfnamefont {B.~I.}\ \bibnamefont {Lundqvist}},\ }\bibfield
  {title} {\bibinfo {title} {Van der {Waals} density functional for general
  geometries},\ }\href@noop {} {\bibfield  {journal} {\bibinfo  {journal}
  {Physical Review Letters}\ }\textbf {\bibinfo {volume} {92}},\ \bibinfo
  {pages} {246401} (\bibinfo {year} {2004})}\BibitemShut {NoStop}%
\bibitem [{\citenamefont {Rom{\'a}n-P{\'e}rez}\ and\ \citenamefont
  {Soler}(2009)}]{roman2009efficient}%
  \BibitemOpen
  \bibfield  {author} {\bibinfo {author} {\bibfnamefont {G.}~\bibnamefont
  {Rom{\'a}n-P{\'e}rez}}\ and\ \bibinfo {author} {\bibfnamefont {J.~M.}\
  \bibnamefont {Soler}},\ }\bibfield  {title} {\bibinfo {title} {Efficient
  implementation of a {Van} der {Waals} density functional: application to
  double-wall carbon nanotubes},\ }\href@noop {} {\bibfield  {journal}
  {\bibinfo  {journal} {Physical Review Letters}\ }\textbf {\bibinfo {volume}
  {103}},\ \bibinfo {pages} {096102} (\bibinfo {year} {2009})}\BibitemShut
  {NoStop}%
\bibitem [{\citenamefont {Klime{\v{s}}}\ \emph {et~al.}(2011)\citenamefont
  {Klime{\v{s}}}, \citenamefont {Bowler},\ and\ \citenamefont
  {Michaelides}}]{klimevs2011van}%
  \BibitemOpen
  \bibfield  {author} {\bibinfo {author} {\bibfnamefont {J.}~\bibnamefont
  {Klime{\v{s}}}}, \bibinfo {author} {\bibfnamefont {D.~R.}\ \bibnamefont
  {Bowler}},\ and\ \bibinfo {author} {\bibfnamefont {A.}~\bibnamefont
  {Michaelides}},\ }\bibfield  {title} {\bibinfo {title} {Van der {Waals}
  density functionals applied to solids},\ }\href@noop {} {\bibfield  {journal}
  {\bibinfo  {journal} {Physical Review B}\ }\textbf {\bibinfo {volume} {83}},\
  \bibinfo {pages} {195131} (\bibinfo {year} {2011})}\BibitemShut {NoStop}%
\bibitem [{\citenamefont {Kashin}\ \emph {et~al.}(2020)\citenamefont {Kashin},
  \citenamefont {Mazurenko}, \citenamefont {Katsnelson},\ and\ \citenamefont
  {Rudenko}}]{kashin2020orbitally}%
  \BibitemOpen
  \bibfield  {author} {\bibinfo {author} {\bibfnamefont {I.}~\bibnamefont
  {Kashin}}, \bibinfo {author} {\bibfnamefont {V.}~\bibnamefont {Mazurenko}},
  \bibinfo {author} {\bibfnamefont {M.}~\bibnamefont {Katsnelson}},\ and\
  \bibinfo {author} {\bibfnamefont {A.}~\bibnamefont {Rudenko}},\ }\bibfield
  {title} {\bibinfo {title} {Orbitally-resolved ferromagnetism of monolayer
  {CrI$_3$}},\ }\href@noop {} {\bibfield  {journal} {\bibinfo  {journal} {2D
  Materials}\ }\textbf {\bibinfo {volume} {7}},\ \bibinfo {pages} {025036}
  (\bibinfo {year} {2020})}\BibitemShut {NoStop}%
\bibitem [{\citenamefont {Haddadi}\ \emph {et~al.}(2024)\citenamefont
  {Haddadi}, \citenamefont {Linscott}, \citenamefont {Timrov}, \citenamefont
  {Marzari},\ and\ \citenamefont {Gibertini}}]{haddadi2023site}%
  \BibitemOpen
  \bibfield  {author} {\bibinfo {author} {\bibfnamefont {F.}~\bibnamefont
  {Haddadi}}, \bibinfo {author} {\bibfnamefont {E.}~\bibnamefont {Linscott}},
  \bibinfo {author} {\bibfnamefont {I.}~\bibnamefont {Timrov}}, \bibinfo
  {author} {\bibfnamefont {N.}~\bibnamefont {Marzari}},\ and\ \bibinfo {author}
  {\bibfnamefont {M.}~\bibnamefont {Gibertini}},\ }\bibfield  {title} {\bibinfo
  {title} {On-site and intersite hubbard corrections in magnetic monolayers:
  The case of {FePS$_3$} and {CrI$_3$}},\ }\href@noop {} {\bibfield  {journal}
  {\bibinfo  {journal} {Physical Review Materials}\ }\textbf {\bibinfo {volume}
  {8}},\ \bibinfo {pages} {014007} (\bibinfo {year} {2024})}\BibitemShut
  {NoStop}%
\bibitem [{\citenamefont {Kumar~Gudelli}\ and\ \citenamefont
  {Guo}(2019)}]{kumar2019magnetism}%
  \BibitemOpen
  \bibfield  {author} {\bibinfo {author} {\bibfnamefont {V.}~\bibnamefont
  {Kumar~Gudelli}}\ and\ \bibinfo {author} {\bibfnamefont {G.-Y.}\ \bibnamefont
  {Guo}},\ }\bibfield  {title} {\bibinfo {title} {Magnetism and magneto-optical
  effects in bulk and few-layer {CrI$_3$}: a theoretical {GGA+U} study},\
  }\href@noop {} {\bibfield  {journal} {\bibinfo  {journal} {New Journal of
  Physics}\ }\textbf {\bibinfo {volume} {21}},\ \bibinfo {pages} {053012}
  (\bibinfo {year} {2019})}\BibitemShut {NoStop}%
\bibitem [{\citenamefont {Kundu}\ \emph {et~al.}(2020)\citenamefont {Kundu},
  \citenamefont {Liu}, \citenamefont {Petrovic},\ and\ \citenamefont
  {Valla}}]{kundu2020valence}%
  \BibitemOpen
  \bibfield  {author} {\bibinfo {author} {\bibfnamefont {A.~K.}\ \bibnamefont
  {Kundu}}, \bibinfo {author} {\bibfnamefont {Y.}~\bibnamefont {Liu}}, \bibinfo
  {author} {\bibfnamefont {C.}~\bibnamefont {Petrovic}},\ and\ \bibinfo
  {author} {\bibfnamefont {T.}~\bibnamefont {Valla}},\ }\bibfield  {title}
  {\bibinfo {title} {Valence band electronic structure of the van der {Waals}
  ferromagnetic insulators: {VI$_3$} and {CrI$_3$}},\ }\href@noop {} {\bibfield
   {journal} {\bibinfo  {journal} {Scientific Reports}\ }\textbf {\bibinfo
  {volume} {10}},\ \bibinfo {pages} {15602} (\bibinfo {year}
  {2020})}\BibitemShut {NoStop}%
\bibitem [{\citenamefont {Hong}\ \emph {et~al.}(2012)\citenamefont {Hong},
  \citenamefont {Stroppa}, \citenamefont {{\'I}niguez}, \citenamefont
  {Picozzi},\ and\ \citenamefont {Vanderbilt}}]{hong2012spin}%
  \BibitemOpen
  \bibfield  {author} {\bibinfo {author} {\bibfnamefont {J.}~\bibnamefont
  {Hong}}, \bibinfo {author} {\bibfnamefont {A.}~\bibnamefont {Stroppa}},
  \bibinfo {author} {\bibfnamefont {J.}~\bibnamefont {{\'I}niguez}}, \bibinfo
  {author} {\bibfnamefont {S.}~\bibnamefont {Picozzi}},\ and\ \bibinfo {author}
  {\bibfnamefont {D.}~\bibnamefont {Vanderbilt}},\ }\bibfield  {title}
  {\bibinfo {title} {Spin-phonon coupling effects in transition-metal
  perovskites: {A DFT+ U} and hybrid-functional study},\ }\href@noop {}
  {\bibfield  {journal} {\bibinfo  {journal} {Physical Review B}\ }\textbf
  {\bibinfo {volume} {85}},\ \bibinfo {pages} {054417} (\bibinfo {year}
  {2012})}\BibitemShut {NoStop}%
\bibitem [{\citenamefont {Li}\ \emph {et~al.}(2013)\citenamefont {Li},
  \citenamefont {Walther}, \citenamefont {Kuc},\ and\ \citenamefont
  {Heine}}]{li2013density}%
  \BibitemOpen
  \bibfield  {author} {\bibinfo {author} {\bibfnamefont {W.}~\bibnamefont
  {Li}}, \bibinfo {author} {\bibfnamefont {C.~F.}\ \bibnamefont {Walther}},
  \bibinfo {author} {\bibfnamefont {A.}~\bibnamefont {Kuc}},\ and\ \bibinfo
  {author} {\bibfnamefont {T.}~\bibnamefont {Heine}},\ }\bibfield  {title}
  {\bibinfo {title} {Density functional theory and beyond for band-gap
  screening: performance for transition-metal oxides and dichalcogenides},\
  }\href@noop {} {\bibfield  {journal} {\bibinfo  {journal} {Journal of
  Chemical Theory and Computation}\ }\textbf {\bibinfo {volume} {9}},\ \bibinfo
  {pages} {2950} (\bibinfo {year} {2013})}\BibitemShut {NoStop}%
\bibitem [{\citenamefont {Huang}\ \emph {et~al.}(2016)\citenamefont {Huang},
  \citenamefont {Ramadugu},\ and\ \citenamefont {Mason}}]{huang2016surface}%
  \BibitemOpen
  \bibfield  {author} {\bibinfo {author} {\bibfnamefont {X.}~\bibnamefont
  {Huang}}, \bibinfo {author} {\bibfnamefont {S.~K.}\ \bibnamefont
  {Ramadugu}},\ and\ \bibinfo {author} {\bibfnamefont {S.~E.}\ \bibnamefont
  {Mason}},\ }\bibfield  {title} {\bibinfo {title} {Surface-specific {DFT+ U}
  approach applied to {$\alpha$-Fe$_2$O$_3$} (0001)},\ }\href@noop {}
  {\bibfield  {journal} {\bibinfo  {journal} {The Journal of Physical Chemistry
  C}\ }\textbf {\bibinfo {volume} {120}},\ \bibinfo {pages} {4919} (\bibinfo
  {year} {2016})}\BibitemShut {NoStop}%
\bibitem [{\citenamefont {Kresse}\ and\ \citenamefont
  {Furthm{\"u}ller}(1996)}]{kresse1996efficient}%
  \BibitemOpen
  \bibfield  {author} {\bibinfo {author} {\bibfnamefont {G.}~\bibnamefont
  {Kresse}}\ and\ \bibinfo {author} {\bibfnamefont {J.}~\bibnamefont
  {Furthm{\"u}ller}},\ }\bibfield  {title} {\bibinfo {title} {Efficient
  iterative schemes for ab initio total-energy calculations using a plane-wave
  basis set},\ }\href@noop {} {\bibfield  {journal} {\bibinfo  {journal}
  {Physical Review B}\ }\textbf {\bibinfo {volume} {54}},\ \bibinfo {pages}
  {11169} (\bibinfo {year} {1996})}\BibitemShut {NoStop}%
\bibitem [{\citenamefont {Perdew}\ \emph {et~al.}(1996)\citenamefont {Perdew},
  \citenamefont {Burke},\ and\ \citenamefont
  {Ernzerhof}}]{perdew1996generalized}%
  \BibitemOpen
  \bibfield  {author} {\bibinfo {author} {\bibfnamefont {J.~P.}\ \bibnamefont
  {Perdew}}, \bibinfo {author} {\bibfnamefont {K.}~\bibnamefont {Burke}},\ and\
  \bibinfo {author} {\bibfnamefont {M.}~\bibnamefont {Ernzerhof}},\ }\bibfield
  {title} {\bibinfo {title} {Generalized gradient approximation made simple},\
  }\href@noop {} {\bibfield  {journal} {\bibinfo  {journal} {Physical Review
  Letters}\ }\textbf {\bibinfo {volume} {77}},\ \bibinfo {pages} {3865}
  (\bibinfo {year} {1996})}\BibitemShut {NoStop}%
\bibitem [{\citenamefont {Wang}\ \emph {et~al.}(2021)\citenamefont {Wang},
  \citenamefont {Xu}, \citenamefont {Liu}, \citenamefont {Tang},\ and\
  \citenamefont {Geng}}]{VASPKIT}%
  \BibitemOpen
  \bibfield  {author} {\bibinfo {author} {\bibfnamefont {V.}~\bibnamefont
  {Wang}}, \bibinfo {author} {\bibfnamefont {N.}~\bibnamefont {Xu}}, \bibinfo
  {author} {\bibfnamefont {J.-C.}\ \bibnamefont {Liu}}, \bibinfo {author}
  {\bibfnamefont {G.}~\bibnamefont {Tang}},\ and\ \bibinfo {author}
  {\bibfnamefont {W.-T.}\ \bibnamefont {Geng}},\ }\bibfield  {title} {\bibinfo
  {title} {{VASPKIT}: A user-friendly interface facilitating high-throughput
  computing and analysis using {VASP} code},\ }\href@noop {} {\bibfield
  {journal} {\bibinfo  {journal} {Computer Physics Communications}\ }\textbf
  {\bibinfo {volume} {267}},\ \bibinfo {pages} {108033} (\bibinfo {year}
  {2021})}\BibitemShut {NoStop}%
\bibitem [{\citenamefont {Wooldridge}(2016)}]{wooldridge2006introduccion}%
  \BibitemOpen
  \bibfield  {author} {\bibinfo {author} {\bibfnamefont {J.~M.}\ \bibnamefont
  {Wooldridge}},\ }\href@noop {} {\emph {\bibinfo {title} {Introductory
  Econometrics: A Modern Approach 6rd ed.}}}\ (\bibinfo  {publisher} {Cengage
  learning},\ \bibinfo {year} {2016})\BibitemShut {NoStop}%
\bibitem [{\citenamefont {Berman}(2016)}]{BERMAN2016135}%
  \BibitemOpen
  \bibfield  {author} {\bibinfo {author} {\bibfnamefont {J.~J.}\ \bibnamefont
  {Berman}},\ }\bibfield  {title} {\bibinfo {title} {Chapter 4 - {Understanding
  Your Data}},\ }in\ \href@noop {} {\emph {\bibinfo {booktitle} {Data
  Simplification}}},\ \bibinfo {editor} {edited by\ \bibinfo {editor}
  {\bibfnamefont {J.~J.}\ \bibnamefont {Berman}}}\ (\bibinfo  {publisher}
  {Morgan Kaufmann},\ \bibinfo {address} {Boston},\ \bibinfo {year} {2016})\
  pp.\ \bibinfo {pages} {135--187}\BibitemShut {NoStop}%
\bibitem [{\citenamefont {Howell}(2010)}]{howell2010log}%
  \BibitemOpen
  \bibfield  {author} {\bibinfo {author} {\bibfnamefont {D.}~\bibnamefont
  {Howell}},\ }\bibfield  {title} {\bibinfo {title} {{Correlation} and
  {Regression}},\ }\href@noop {} {\bibfield  {journal} {\bibinfo  {journal}
  {Statistical Methods for Psychology. 7th Edition Canada, Wadsworth CENGAGE
  Learning}\ } (\bibinfo {year} {2010})}\BibitemShut {NoStop}%
\bibitem [{\citenamefont {Nisar}\ \emph {et~al.}(2011)\citenamefont {Nisar},
  \citenamefont {{\AA}rhammar}, \citenamefont {J{\"a}mstorp},\ and\
  \citenamefont {Ahuja}}]{nisar2011optical}%
  \BibitemOpen
  \bibfield  {author} {\bibinfo {author} {\bibfnamefont {J.}~\bibnamefont
  {Nisar}}, \bibinfo {author} {\bibfnamefont {C.}~\bibnamefont {{\AA}rhammar}},
  \bibinfo {author} {\bibfnamefont {E.}~\bibnamefont {J{\"a}mstorp}},\ and\
  \bibinfo {author} {\bibfnamefont {R.}~\bibnamefont {Ahuja}},\ }\bibfield
  {title} {\bibinfo {title} {Optical gap and native point defects in kaolinite
  studied by the {GGA-PBE}, {HSE} functional, and {GW} approaches},\
  }\href@noop {} {\bibfield  {journal} {\bibinfo  {journal} {Physical Review
  B}\ }\textbf {\bibinfo {volume} {84}},\ \bibinfo {pages} {075120} (\bibinfo
  {year} {2011})}\BibitemShut {NoStop}%
\bibitem [{\citenamefont {Jiang}\ \emph {et~al.}(2018)\citenamefont {Jiang},
  \citenamefont {Li}, \citenamefont {Liao}, \citenamefont {Zhao},\ and\
  \citenamefont {Zhong}}]{jiang2018spin}%
  \BibitemOpen
  \bibfield  {author} {\bibinfo {author} {\bibfnamefont {P.}~\bibnamefont
  {Jiang}}, \bibinfo {author} {\bibfnamefont {L.}~\bibnamefont {Li}}, \bibinfo
  {author} {\bibfnamefont {Z.}~\bibnamefont {Liao}}, \bibinfo {author}
  {\bibfnamefont {Y.}~\bibnamefont {Zhao}},\ and\ \bibinfo {author}
  {\bibfnamefont {Z.}~\bibnamefont {Zhong}},\ }\bibfield  {title} {\bibinfo
  {title} {Spin direction-controlled electronic band structure in
  two-dimensional ferromagnetic {CrI$_3$}},\ }\href@noop {} {\bibfield
  {journal} {\bibinfo  {journal} {Nano Letters}\ }\textbf {\bibinfo {volume}
  {18}},\ \bibinfo {pages} {3844} (\bibinfo {year} {2018})}\BibitemShut
  {NoStop}%
\bibitem [{\citenamefont {Zhang}\ \emph {et~al.}(2018)\citenamefont {Zhang},
  \citenamefont {Zhao}, \citenamefont {Zhou}, \citenamefont {Xue},
  \citenamefont {Ma},\ and\ \citenamefont {Yang}}]{zhang2018strong}%
  \BibitemOpen
  \bibfield  {author} {\bibinfo {author} {\bibfnamefont {J.}~\bibnamefont
  {Zhang}}, \bibinfo {author} {\bibfnamefont {B.}~\bibnamefont {Zhao}},
  \bibinfo {author} {\bibfnamefont {T.}~\bibnamefont {Zhou}}, \bibinfo {author}
  {\bibfnamefont {Y.}~\bibnamefont {Xue}}, \bibinfo {author} {\bibfnamefont
  {C.}~\bibnamefont {Ma}},\ and\ \bibinfo {author} {\bibfnamefont
  {Z.}~\bibnamefont {Yang}},\ }\bibfield  {title} {\bibinfo {title} {Strong
  magnetization and chern insulators in compressed graphene/{CrI$_3$} {Van} der
  {Waals} heterostructures},\ }\href@noop {} {\bibfield  {journal} {\bibinfo
  {journal} {Physical Review B}\ }\textbf {\bibinfo {volume} {97}},\ \bibinfo
  {pages} {085401} (\bibinfo {year} {2018})}\BibitemShut {NoStop}%
\bibitem [{\citenamefont {Jiao}\ \emph {et~al.}(2022)\citenamefont {Jiao},
  \citenamefont {Fu},\ and\ \citenamefont {Huang}}]{jiao2022descriptors}%
  \BibitemOpen
  \bibfield  {author} {\bibinfo {author} {\bibfnamefont {S.}~\bibnamefont
  {Jiao}}, \bibinfo {author} {\bibfnamefont {X.}~\bibnamefont {Fu}},\ and\
  \bibinfo {author} {\bibfnamefont {H.}~\bibnamefont {Huang}},\ }\bibfield
  {title} {\bibinfo {title} {Descriptors for the evaluation of electrocatalytic
  reactions: d-band theory and beyond},\ }\href@noop {} {\bibfield  {journal}
  {\bibinfo  {journal} {Advanced Functional Materials}\ }\textbf {\bibinfo
  {volume} {32}},\ \bibinfo {pages} {2107651} (\bibinfo {year}
  {2022})}\BibitemShut {NoStop}%
\bibitem [{\citenamefont {Zhu}\ \emph {et~al.}(2022)\citenamefont {Zhu},
  \citenamefont {Huang}, \citenamefont {Huang}, \citenamefont {Gao},
  \citenamefont {Su},\ and\ \citenamefont {Qiao}}]{zhu2022d}%
  \BibitemOpen
  \bibfield  {author} {\bibinfo {author} {\bibfnamefont {Q.}~\bibnamefont
  {Zhu}}, \bibinfo {author} {\bibfnamefont {W.}~\bibnamefont {Huang}}, \bibinfo
  {author} {\bibfnamefont {C.}~\bibnamefont {Huang}}, \bibinfo {author}
  {\bibfnamefont {L.}~\bibnamefont {Gao}}, \bibinfo {author} {\bibfnamefont
  {Y.}~\bibnamefont {Su}},\ and\ \bibinfo {author} {\bibfnamefont
  {L.}~\bibnamefont {Qiao}},\ }\bibfield  {title} {\bibinfo {title} {The d band
  center as an indicator for the hydrogen solution and diffusion behaviors in
  transition metals},\ }\href@noop {} {\bibfield  {journal} {\bibinfo
  {journal} {International Journal of Hydrogen Energy}\ }\textbf {\bibinfo
  {volume} {47}},\ \bibinfo {pages} {38445} (\bibinfo {year}
  {2022})}\BibitemShut {NoStop}%
\bibitem [{\citenamefont {Demiroglu}\ \emph {et~al.}(2016)\citenamefont
  {Demiroglu}, \citenamefont {Li}, \citenamefont {Piccolo},\ and\ \citenamefont
  {Johnston}}]{demiroglu2016dft}%
  \BibitemOpen
  \bibfield  {author} {\bibinfo {author} {\bibfnamefont {I.}~\bibnamefont
  {Demiroglu}}, \bibinfo {author} {\bibfnamefont {Z.}~\bibnamefont {Li}},
  \bibinfo {author} {\bibfnamefont {L.}~\bibnamefont {Piccolo}},\ and\ \bibinfo
  {author} {\bibfnamefont {R.~L.}\ \bibnamefont {Johnston}},\ }\bibfield
  {title} {\bibinfo {title} {A {DFT} study of molecular adsorption on {Au--Rh}
  nanoalloys},\ }\href@noop {} {\bibfield  {journal} {\bibinfo  {journal}
  {Catalysis Science \& Technology}\ }\textbf {\bibinfo {volume} {6}},\
  \bibinfo {pages} {6916} (\bibinfo {year} {2016})}\BibitemShut {NoStop}%
\bibitem [{\citenamefont {Guo}\ \emph {et~al.}(2018)\citenamefont {Guo},
  \citenamefont {Yuan}, \citenamefont {Wang}, \citenamefont {Shi},\ and\
  \citenamefont {Wang}}]{guo2018half}%
  \BibitemOpen
  \bibfield  {author} {\bibinfo {author} {\bibfnamefont {Y.}~\bibnamefont
  {Guo}}, \bibinfo {author} {\bibfnamefont {S.}~\bibnamefont {Yuan}}, \bibinfo
  {author} {\bibfnamefont {B.}~\bibnamefont {Wang}}, \bibinfo {author}
  {\bibfnamefont {L.}~\bibnamefont {Shi}},\ and\ \bibinfo {author}
  {\bibfnamefont {J.}~\bibnamefont {Wang}},\ }\bibfield  {title} {\bibinfo
  {title} {Half-metallicity and enhanced ferromagnetism in {Li}-adsorbed
  ultrathin chromium triiodide},\ }\href@noop {} {\bibfield  {journal}
  {\bibinfo  {journal} {Journal of Materials Chemistry C}\ }\textbf {\bibinfo
  {volume} {6}},\ \bibinfo {pages} {5716} (\bibinfo {year} {2018})}\BibitemShut
  {NoStop}%
\bibitem [{\citenamefont {Wang}\ \emph {et~al.}(2016)\citenamefont {Wang},
  \citenamefont {Fan}, \citenamefont {Zhu},\ and\ \citenamefont
  {Wu}}]{wang2016doping}%
  \BibitemOpen
  \bibfield  {author} {\bibinfo {author} {\bibfnamefont {H.}~\bibnamefont
  {Wang}}, \bibinfo {author} {\bibfnamefont {F.}~\bibnamefont {Fan}}, \bibinfo
  {author} {\bibfnamefont {S.}~\bibnamefont {Zhu}},\ and\ \bibinfo {author}
  {\bibfnamefont {H.}~\bibnamefont {Wu}},\ }\bibfield  {title} {\bibinfo
  {title} {Doping enhanced ferromagnetism and induced half-metallicity in
  {CrI$_3$} monolayer},\ }\href@noop {} {\bibfield  {journal} {\bibinfo
  {journal} {Europhysics Letters}\ }\textbf {\bibinfo {volume} {114}},\
  \bibinfo {pages} {47001} (\bibinfo {year} {2016})}\BibitemShut {NoStop}%
\bibitem [{\citenamefont {Webster}\ and\ \citenamefont
  {Yan}(2018)}]{webster2018strain}%
  \BibitemOpen
  \bibfield  {author} {\bibinfo {author} {\bibfnamefont {L.}~\bibnamefont
  {Webster}}\ and\ \bibinfo {author} {\bibfnamefont {J.-A.}\ \bibnamefont
  {Yan}},\ }\bibfield  {title} {\bibinfo {title} {Strain-tunable magnetic
  anisotropy in monolayer {CrCl$_3$}, {CrBr$_3$}, and {CrI$_3$}},\ }\href@noop
  {} {\bibfield  {journal} {\bibinfo  {journal} {Physical Review B}\ }\textbf
  {\bibinfo {volume} {98}},\ \bibinfo {pages} {144411} (\bibinfo {year}
  {2018})}\BibitemShut {NoStop}%
\bibitem [{\citenamefont {Ghosh}\ \emph {et~al.}(2019)\citenamefont {Ghosh},
  \citenamefont {Stoji{\'c}},\ and\ \citenamefont
  {Binggeli}}]{ghosh2019structural}%
  \BibitemOpen
  \bibfield  {author} {\bibinfo {author} {\bibfnamefont {S.}~\bibnamefont
  {Ghosh}}, \bibinfo {author} {\bibfnamefont {N.}~\bibnamefont {Stoji{\'c}}},\
  and\ \bibinfo {author} {\bibfnamefont {N.}~\bibnamefont {Binggeli}},\
  }\bibfield  {title} {\bibinfo {title} {Structural and magnetic response of
  {CrI$_3$} monolayer to electric field},\ }\href@noop {} {\bibfield  {journal}
  {\bibinfo  {journal} {Physica B: Condensed Matter}\ }\textbf {\bibinfo
  {volume} {570}},\ \bibinfo {pages} {166} (\bibinfo {year}
  {2019})}\BibitemShut {NoStop}%
\bibitem [{\citenamefont {Campo}\ and\ \citenamefont
  {Cococcioni}(2010)}]{campo2010extended}%
  \BibitemOpen
  \bibfield  {author} {\bibinfo {author} {\bibfnamefont {V.~L.}\ \bibnamefont
  {Campo}}\ and\ \bibinfo {author} {\bibfnamefont {M.}~\bibnamefont
  {Cococcioni}},\ }\bibfield  {title} {\bibinfo {title} {Extended {DFT+U+V}
  method with on-site and inter-site electronic interactions},\ }\href@noop {}
  {\bibfield  {journal} {\bibinfo  {journal} {Journal of Physics: Condensed
  Matter}\ }\textbf {\bibinfo {volume} {22}},\ \bibinfo {pages} {055602}
  (\bibinfo {year} {2010})}\BibitemShut {NoStop}%
\bibitem [{\citenamefont {Yu}\ \emph {et~al.}(2023)\citenamefont {Yu},
  \citenamefont {Zhang}, \citenamefont {Wan}, \citenamefont {Guo},
  \citenamefont {Gui}, \citenamefont {Peng}, \citenamefont {Li}, \citenamefont
  {Fu}, \citenamefont {Lu}, \citenamefont {Ye} \emph {et~al.}}]{yu2023active}%
  \BibitemOpen
  \bibfield  {author} {\bibinfo {author} {\bibfnamefont {W.}~\bibnamefont
  {Yu}}, \bibinfo {author} {\bibfnamefont {Z.}~\bibnamefont {Zhang}}, \bibinfo
  {author} {\bibfnamefont {X.}~\bibnamefont {Wan}}, \bibinfo {author}
  {\bibfnamefont {H.}~\bibnamefont {Guo}}, \bibinfo {author} {\bibfnamefont
  {Q.}~\bibnamefont {Gui}}, \bibinfo {author} {\bibfnamefont {Y.}~\bibnamefont
  {Peng}}, \bibinfo {author} {\bibfnamefont {Y.}~\bibnamefont {Li}}, \bibinfo
  {author} {\bibfnamefont {W.}~\bibnamefont {Fu}}, \bibinfo {author}
  {\bibfnamefont {D.}~\bibnamefont {Lu}}, \bibinfo {author} {\bibfnamefont
  {Y.}~\bibnamefont {Ye}}, \emph {et~al.},\ }\bibfield  {title} {\bibinfo
  {title} {Active learning the high-dimensional transferable {Hubbard} {U} and
  {V} parameters in the {DFT+ U+ V} scheme},\ }\href@noop {} {\bibfield
  {journal} {\bibinfo  {journal} {Journal of Chemical Theory and Computation}\
  }\textbf {\bibinfo {volume} {19}},\ \bibinfo {pages} {6425} (\bibinfo {year}
  {2023})}\BibitemShut {NoStop}%
\bibitem [{\citenamefont {Liu}\ \emph {et~al.}(2018)\citenamefont {Liu},
  \citenamefont {Shi}, \citenamefont {Lu},\ and\ \citenamefont
  {Anantram}}]{liu2018analysis}%
  \BibitemOpen
  \bibfield  {author} {\bibinfo {author} {\bibfnamefont {J.}~\bibnamefont
  {Liu}}, \bibinfo {author} {\bibfnamefont {M.}~\bibnamefont {Shi}}, \bibinfo
  {author} {\bibfnamefont {J.}~\bibnamefont {Lu}},\ and\ \bibinfo {author}
  {\bibfnamefont {M.}~\bibnamefont {Anantram}},\ }\bibfield  {title} {\bibinfo
  {title} {Analysis of electrical-field-dependent {Dzyaloshinskii-Moriya}
  interaction and magnetocrystalline anisotropy in a two-dimensional
  ferromagnetic monolayer},\ }\href@noop {} {\bibfield  {journal} {\bibinfo
  {journal} {Physical Review B}\ }\textbf {\bibinfo {volume} {97}},\ \bibinfo
  {pages} {054416} (\bibinfo {year} {2018})}\BibitemShut {NoStop}%
\bibitem [{\citenamefont {Le{\'o}n}\ \emph {et~al.}(2020)\citenamefont
  {Le{\'o}n}, \citenamefont {Gonz{\'a}lez}, \citenamefont
  {Mej{\'\i}a-L{\'o}pez}, \citenamefont {de~Lima},\ and\ \citenamefont
  {Morell}}]{leon2020strain}%
  \BibitemOpen
  \bibfield  {author} {\bibinfo {author} {\bibfnamefont {A.~M.}\ \bibnamefont
  {Le{\'o}n}}, \bibinfo {author} {\bibfnamefont {J.~W.}\ \bibnamefont
  {Gonz{\'a}lez}}, \bibinfo {author} {\bibfnamefont {J.}~\bibnamefont
  {Mej{\'\i}a-L{\'o}pez}}, \bibinfo {author} {\bibfnamefont {F.~C.}\
  \bibnamefont {de~Lima}},\ and\ \bibinfo {author} {\bibfnamefont {E.~S.}\
  \bibnamefont {Morell}},\ }\bibfield  {title} {\bibinfo {title}
  {Strain-induced phase transition in \ce{CrI3} bilayers},\ }\href@noop {}
  {\bibfield  {journal} {\bibinfo  {journal} {2D Materials}\ }\textbf {\bibinfo
  {volume} {7}},\ \bibinfo {pages} {035008} (\bibinfo {year}
  {2020})}\BibitemShut {NoStop}%
\bibitem [{\citenamefont {Morell}\ \emph {et~al.}(2019)\citenamefont {Morell},
  \citenamefont {Le{\'o}n}, \citenamefont {Miwa},\ and\ \citenamefont
  {Vargas}}]{morell2019control}%
  \BibitemOpen
  \bibfield  {author} {\bibinfo {author} {\bibfnamefont {E.~S.}\ \bibnamefont
  {Morell}}, \bibinfo {author} {\bibfnamefont {A.}~\bibnamefont {Le{\'o}n}},
  \bibinfo {author} {\bibfnamefont {R.~H.}\ \bibnamefont {Miwa}},\ and\
  \bibinfo {author} {\bibfnamefont {P.}~\bibnamefont {Vargas}},\ }\bibfield
  {title} {\bibinfo {title} {Control of magnetism in bilayer {CrI$_3$} by an
  external electric field},\ }\href@noop {} {\bibfield  {journal} {\bibinfo
  {journal} {2D Materials}\ }\textbf {\bibinfo {volume} {6}},\ \bibinfo {pages}
  {025020} (\bibinfo {year} {2019})}\BibitemShut {NoStop}%
\bibitem [{\citenamefont {Huang}\ \emph {et~al.}(2018)\citenamefont {Huang},
  \citenamefont {Clark}, \citenamefont {Klein}, \citenamefont {MacNeill},
  \citenamefont {Navarro-Moratalla}, \citenamefont {Seyler}, \citenamefont
  {Wilson}, \citenamefont {McGuire}, \citenamefont {Cobden}, \citenamefont
  {Xiao} \emph {et~al.}}]{huang2018electrical}%
  \BibitemOpen
  \bibfield  {author} {\bibinfo {author} {\bibfnamefont {B.}~\bibnamefont
  {Huang}}, \bibinfo {author} {\bibfnamefont {G.}~\bibnamefont {Clark}},
  \bibinfo {author} {\bibfnamefont {D.~R.}\ \bibnamefont {Klein}}, \bibinfo
  {author} {\bibfnamefont {D.}~\bibnamefont {MacNeill}}, \bibinfo {author}
  {\bibfnamefont {E.}~\bibnamefont {Navarro-Moratalla}}, \bibinfo {author}
  {\bibfnamefont {K.~L.}\ \bibnamefont {Seyler}}, \bibinfo {author}
  {\bibfnamefont {N.}~\bibnamefont {Wilson}}, \bibinfo {author} {\bibfnamefont
  {M.~A.}\ \bibnamefont {McGuire}}, \bibinfo {author} {\bibfnamefont {D.~H.}\
  \bibnamefont {Cobden}}, \bibinfo {author} {\bibfnamefont {D.}~\bibnamefont
  {Xiao}}, \emph {et~al.},\ }\bibfield  {title} {\bibinfo {title} {Electrical
  control of {2D} magnetism in bilayer {CrI$_3$}},\ }\href@noop {} {\bibfield
  {journal} {\bibinfo  {journal} {Nature Nanotechnology}\ }\textbf {\bibinfo
  {volume} {13}},\ \bibinfo {pages} {544} (\bibinfo {year} {2018})}\BibitemShut
  {NoStop}%
\bibitem [{\citenamefont {Thonhauser}\ \emph {et~al.}(2007)\citenamefont
  {Thonhauser}, \citenamefont {Cooper}, \citenamefont {Li}, \citenamefont
  {Puzder}, \citenamefont {Hyldgaard},\ and\ \citenamefont
  {Langreth}}]{thonhauser2007van}%
  \BibitemOpen
  \bibfield  {author} {\bibinfo {author} {\bibfnamefont {T.}~\bibnamefont
  {Thonhauser}}, \bibinfo {author} {\bibfnamefont {V.~R.}\ \bibnamefont
  {Cooper}}, \bibinfo {author} {\bibfnamefont {S.}~\bibnamefont {Li}}, \bibinfo
  {author} {\bibfnamefont {A.}~\bibnamefont {Puzder}}, \bibinfo {author}
  {\bibfnamefont {P.}~\bibnamefont {Hyldgaard}},\ and\ \bibinfo {author}
  {\bibfnamefont {D.~C.}\ \bibnamefont {Langreth}},\ }\bibfield  {title}
  {\bibinfo {title} {{Van} der {Waals} density functional: Self-consistent
  potential and the nature of the {Van} der {Waals} bond},\ }\href@noop {}
  {\bibfield  {journal} {\bibinfo  {journal} {Physical Review B}\ }\textbf
  {\bibinfo {volume} {76}},\ \bibinfo {pages} {125112} (\bibinfo {year}
  {2007})}\BibitemShut {NoStop}%
\bibitem [{Non()}]{Nonlocal}%
  \BibitemOpen
  \href {https://www.vasp.at/wiki/index.php/Nonlocal_vdW-DF_functionals}
  {\bibinfo {title} {Nonlocal {vdW-DF} functionals - {VASP Wiki}}},\ \bibinfo
  {howpublished}
  {\url{https://www.vasp.at/wiki/index.php/Nonlocal_vdW-DF_functionals}}\BibitemShut
  {NoStop}%
\bibitem [{\citenamefont {Jiang}\ \emph {et~al.}(2019)\citenamefont {Jiang},
  \citenamefont {Wang}, \citenamefont {Chen}, \citenamefont {Zhong},
  \citenamefont {Yuan}, \citenamefont {Lu},\ and\ \citenamefont
  {Ji}}]{jiang2019stacking}%
  \BibitemOpen
  \bibfield  {author} {\bibinfo {author} {\bibfnamefont {P.}~\bibnamefont
  {Jiang}}, \bibinfo {author} {\bibfnamefont {C.}~\bibnamefont {Wang}},
  \bibinfo {author} {\bibfnamefont {D.}~\bibnamefont {Chen}}, \bibinfo {author}
  {\bibfnamefont {Z.}~\bibnamefont {Zhong}}, \bibinfo {author} {\bibfnamefont
  {Z.}~\bibnamefont {Yuan}}, \bibinfo {author} {\bibfnamefont {Z.-Y.}\
  \bibnamefont {Lu}},\ and\ \bibinfo {author} {\bibfnamefont {W.}~\bibnamefont
  {Ji}},\ }\bibfield  {title} {\bibinfo {title} {Stacking tunable interlayer
  magnetism in bilayer {CrI$_3$}},\ }\href@noop {} {\bibfield  {journal}
  {\bibinfo  {journal} {Physical Review B}\ }\textbf {\bibinfo {volume} {99}},\
  \bibinfo {pages} {144401} (\bibinfo {year} {2019})}\BibitemShut {NoStop}%
\bibitem [{\citenamefont {Guo}\ \emph {et~al.}(2020)\citenamefont {Guo},
  \citenamefont {Liu}, \citenamefont {Zhao}, \citenamefont {Jiang},
  \citenamefont {Zhou},\ and\ \citenamefont {Zhao}}]{guo2020enhanced}%
  \BibitemOpen
  \bibfield  {author} {\bibinfo {author} {\bibfnamefont {Y.}~\bibnamefont
  {Guo}}, \bibinfo {author} {\bibfnamefont {N.}~\bibnamefont {Liu}}, \bibinfo
  {author} {\bibfnamefont {Y.}~\bibnamefont {Zhao}}, \bibinfo {author}
  {\bibfnamefont {X.}~\bibnamefont {Jiang}}, \bibinfo {author} {\bibfnamefont
  {S.}~\bibnamefont {Zhou}},\ and\ \bibinfo {author} {\bibfnamefont
  {J.}~\bibnamefont {Zhao}},\ }\bibfield  {title} {\bibinfo {title} {Enhanced
  ferromagnetism of {CrI$_3$} bilayer by self-intercalation},\ }\href@noop {}
  {\bibfield  {journal} {\bibinfo  {journal} {Chinese Physics Letters}\
  }\textbf {\bibinfo {volume} {37}},\ \bibinfo {pages} {107506} (\bibinfo
  {year} {2020})}\BibitemShut {NoStop}%
\end{thebibliography}


\pagebreak
\newpage\null\thispagestyle{empty}\newpage
\widetext

\begin{center}
\textbf{\large Supplemental Material:  \\Fine-Tuning Magnetism in \ce{CrI3} Monolayers and Bilayers: A DFT+U Approach }

Diego Lauer$^{1,2}$,  Jhon W. Gonz\'{a}lez$^3$, Eric Su\'{a}rez Morell$^1$, and Andr\'es Ayuela$^2$\\
$^1$ Departamento de F\'{i}sica, Universidad T\'{e}cnica Federico Santa Mar\'{i}a, Casilla Postal 110V, Valpara\'{i}so, Chile.\\
$^2$ Centro de F\'{i}sica de Materiales-Materials Physics Center CFM-MPC CSIC-UPV/EHU,
Donostia International Physics Center, Paseo Manuel Lardizabal 5, Donostia–San Sebasti\'{a}n 20018, Spain\\
$^3$ Departamento de F\'{i}sica, Universidad de Antofagasta,
Avenida Angamos 601, Casilla 170, Antofagasta, Chile\\

\end{center}
%\begin{widetext}

%\end{widetext}
%%%%% SM
\setcounter{figure}{0} 
\setcounter{section}{0} 
\setcounter{equation}{0}
\setcounter{page}{1}
\renewcommand{\thepage}{S\arabic{page}} 
\renewcommand{\thesection}{S\Roman{section}}   
\renewcommand{\thetable}{S\arabic{table}}  
\renewcommand{\thefigure}{S\arabic{figure}} 
\renewcommand{\theequation}{S\arabic{equation}} 
%\renewcommand{\bibnumfmt}[1]{[S#1]}
%\renewcommand{ \citenumfont}[1]{S#1}

%\section{titulo }

\begin{figure}[h]
\includegraphics[clip,width=0.4\textwidth,clip]{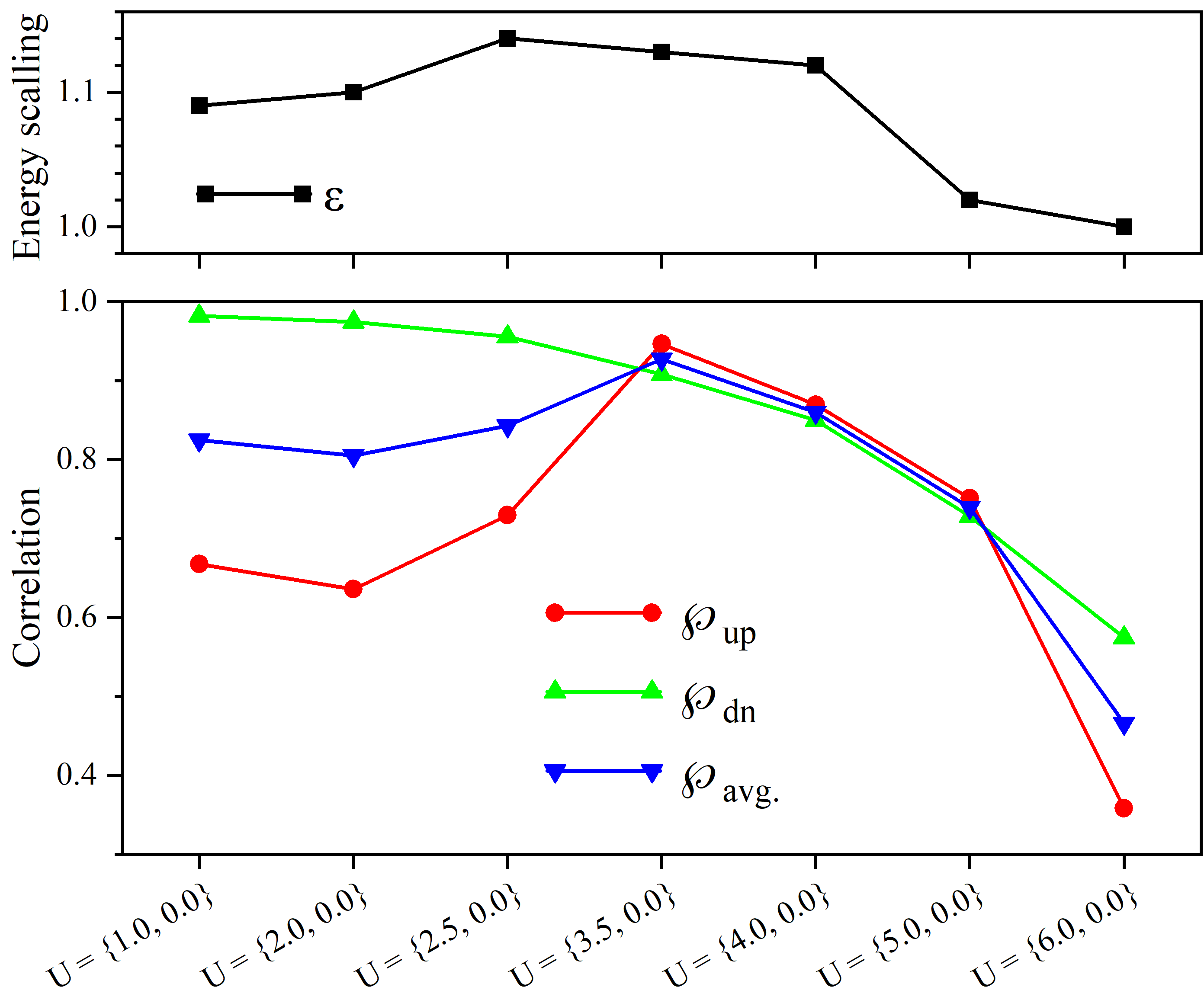}
\includegraphics[clip,width=0.41\textwidth,clip]{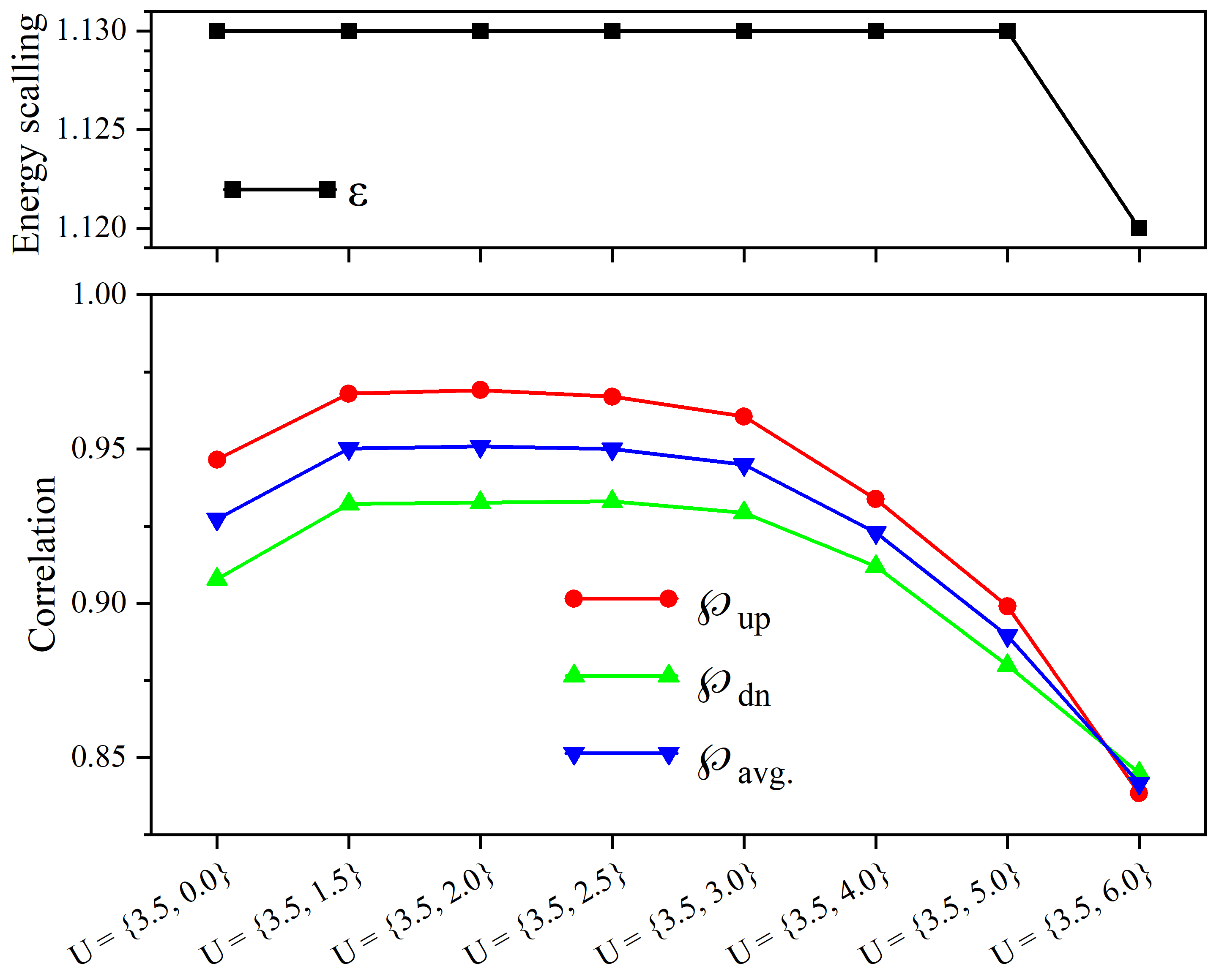}
\caption{Energy scaling and Pearson Correlation $\mathcal{P}$ as a function of different $U = \left\lbrace U_{Cr}, U_I \right\rbrace $.  The values $\mathcal{P}$ 
for spin-up, spin-down, and the average between both are showed in the bottom panels. The left panels correspond to the variation of $U_{Cr}$ for $U_I =0$, and the right panels correspond to the variation of $U_I$ for a fixed $U_{Cr} = 3.5$ eV.}
\label{fig:scalling}
\end{figure}

%%%

\begin{figure}[h!]
\includegraphics[clip,width=.5\textwidth,angle=0,clip]{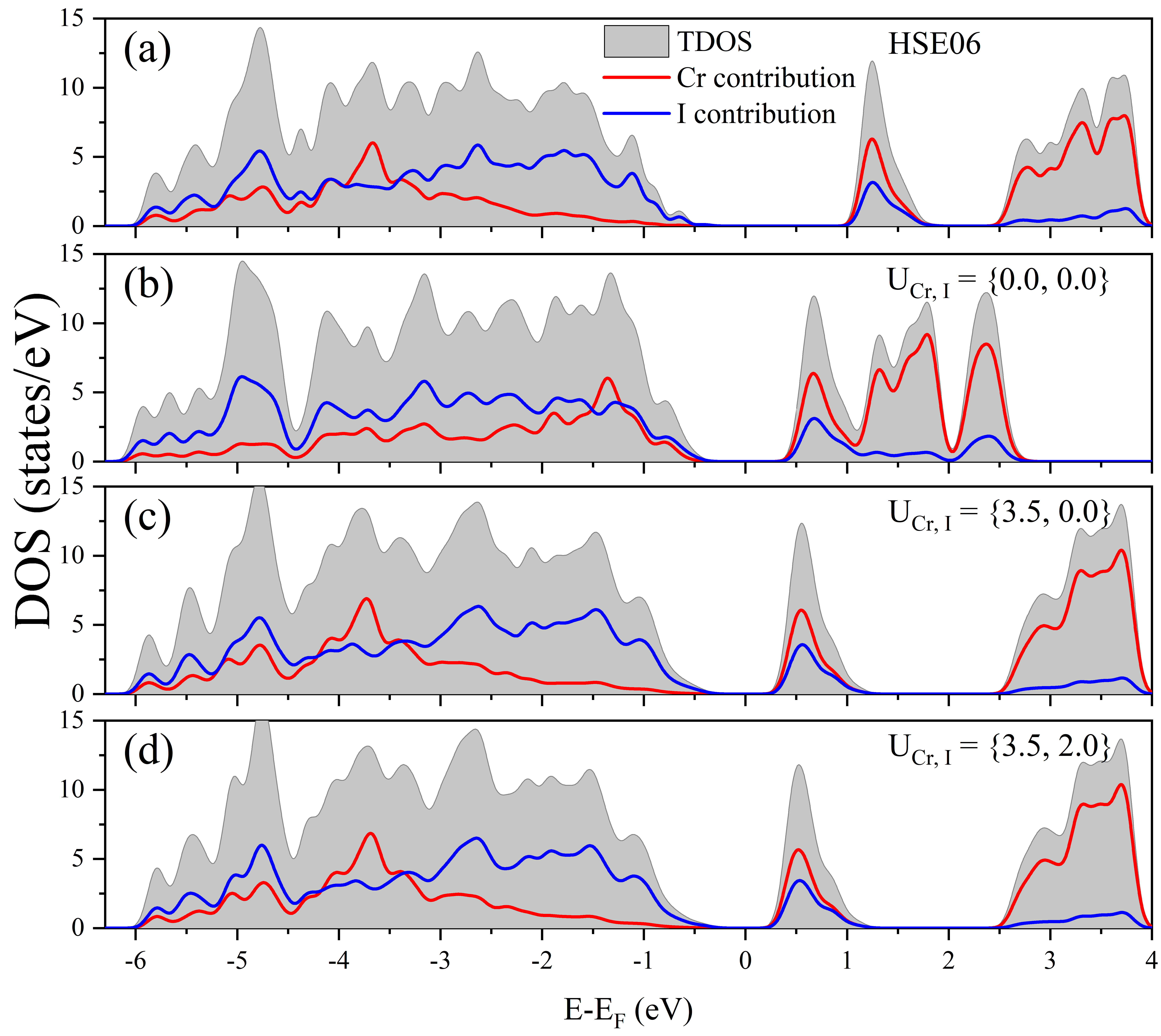}
    \caption{Density of states for several parameters with spin-orbit coupling.  
    The color scale and the plot distribution follows the caption of Fig. \ref{fig:DOS-por-atomo}. In panel (a), results for HSE06 functional, in (b) $U_{Cr,I}=\left\lbrace 0.0\,,0.0 \right\rbrace$, in (c) $U_{Cr,I}=\left\lbrace 3.5\,,0.0 \right\rbrace$, and (d) $U_{Cr,I}=\left\lbrace 3.5\,,2.0 \right\rbrace$.}
    \label{fig:DOS-por-atomo-SOC}
\end{figure}

\begin{figure}[ht!]
\includegraphics[clip,width=0.5\columnwidth,angle=0,clip]{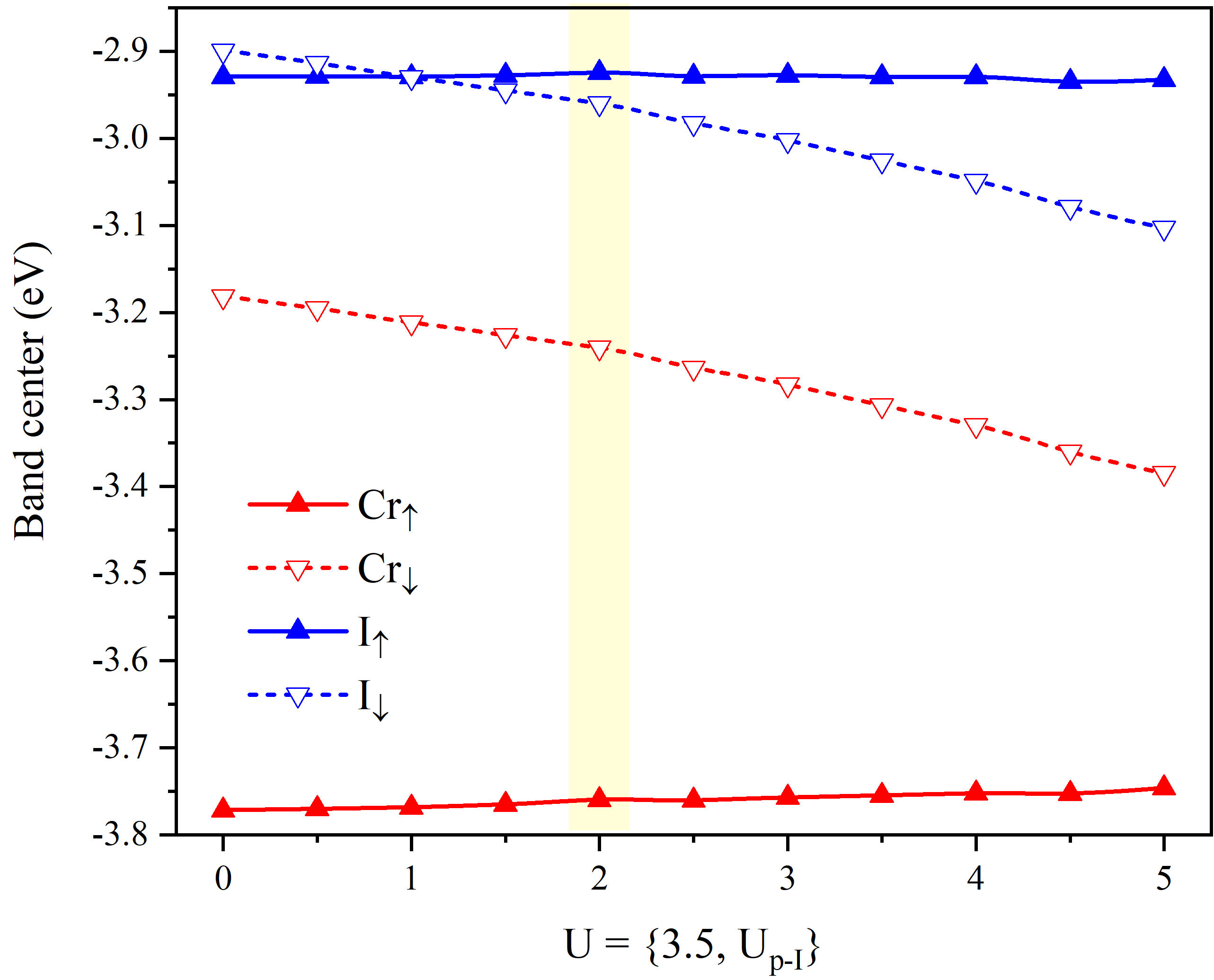}
\caption{Band center as a function of the Hubbard U parameter with scaling applied. The red line represents the d-band center for chromium (Cr) in spin-up ($\uparrow$) and spin-down ($\downarrow$) states, while the blue line represent the p-band center for iodine (I) in spin-up ($\uparrow$) and spin-down ($\downarrow$) states. The panel shows the U(I) variation in the range [0.0, 5.5] eV with U(Cr) set to $3.5$ eV.}
\label{fig:Band-CENTER-escalado}
\end{figure}

\begin{figure}[ht!]
\includegraphics[clip,width=.5\columnwidth,angle=0,clip]{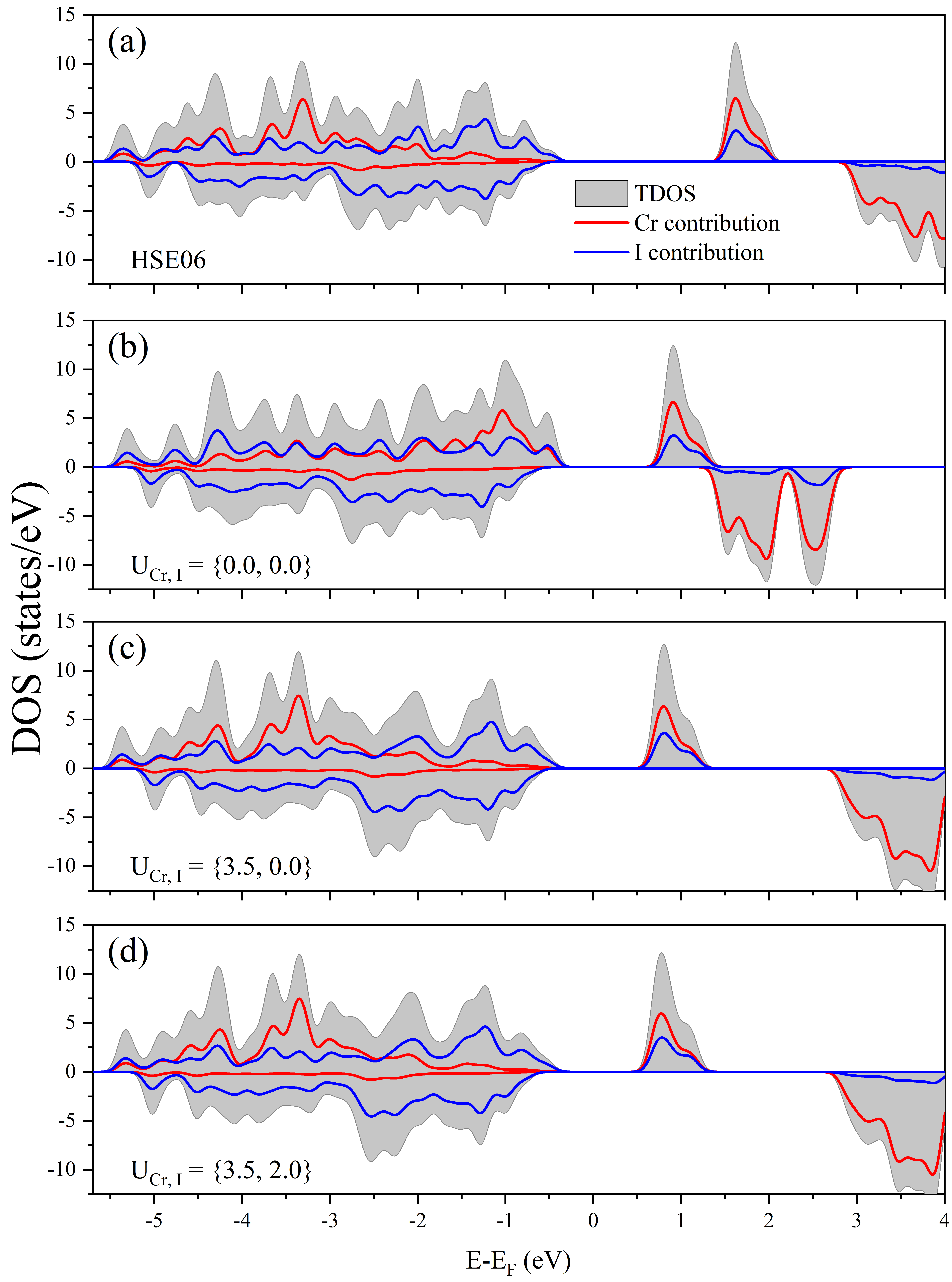}
\caption{Density of states projected by atomic species is calculated using different approaches. 
The total density of states is divided in the sum over the d-orbitals of chromium and the sum over the p-orbitals of iodine using (a)  HSE06 hybrid functional, (b) 
$U_{Cr, I}=\left\lbrace 0.0,\,0.0 \right\rbrace$, (c) $U_{Cr, I}=\left\lbrace 3.5,\,0.0 \right\rbrace$, 
and (d)  $U_{Cr, I}=\left\lbrace 3.5,\,2.0 \right\rbrace$.}
\label{fig:DOS-por-atomo}
\end{figure}

\begin{figure}[h]
\includegraphics[clip,width=0.5\textwidth,clip]{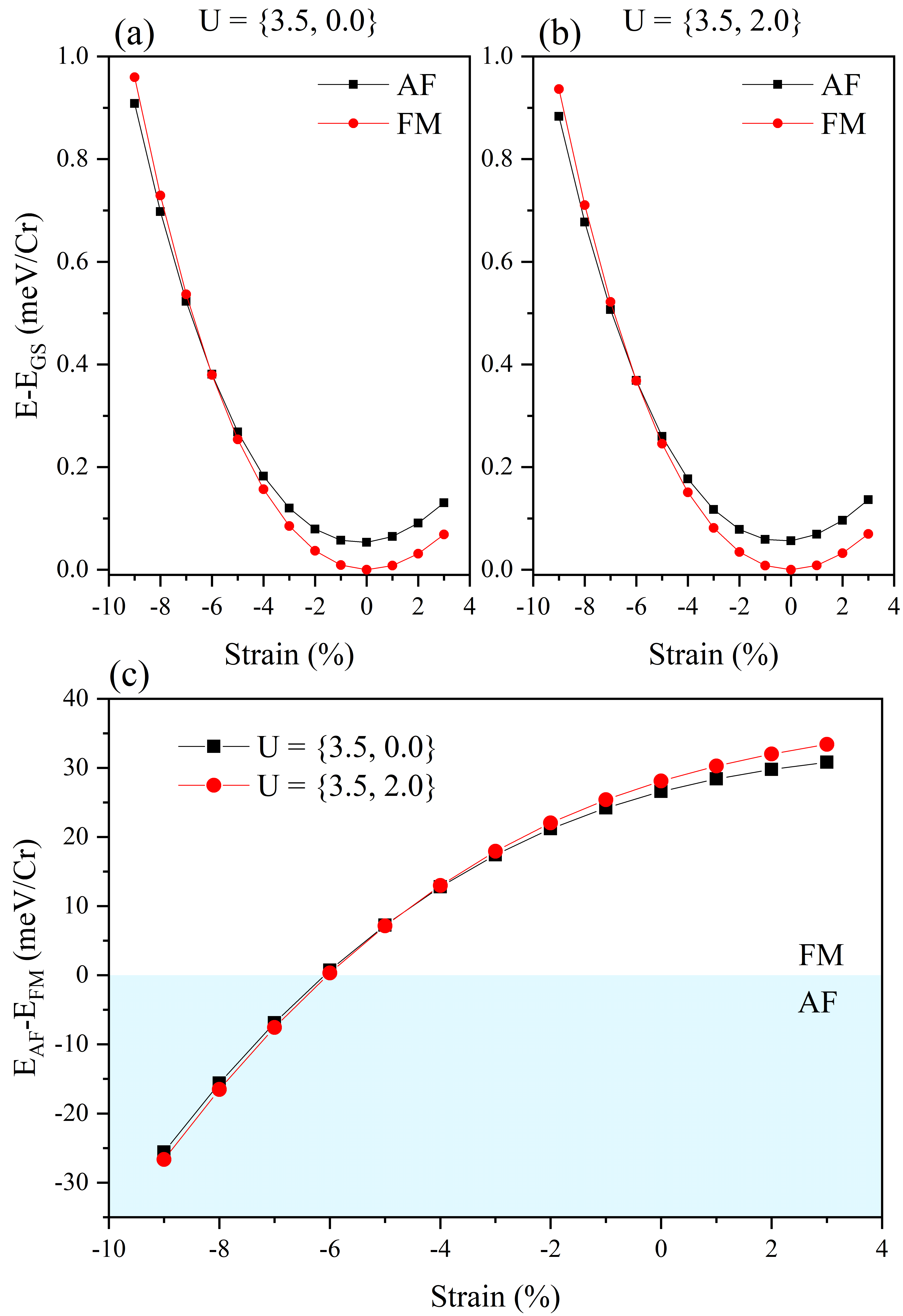}
    \caption{Total energy as a function of strain for the two magnetic ordering in a monolayer of \ce{CrI3}. (a) $U_{Cr,I}=\left\lbrace 3.5,\,0.0 \right\rbrace$, and (b) $U_{Cr,I}=\left\lbrace 3.5,\,2.0 \right\rbrace$ configuration. The reference energy E$_{GS}$ is the energy of the FM case with $0\%$. The (c) pannel shows the AF-FM energy difference for both U cases.}
    \label{fig:strain-monolayer}
\end{figure}

\begin{figure}[h]
\includegraphics[clip,width=0.5\textwidth,clip]{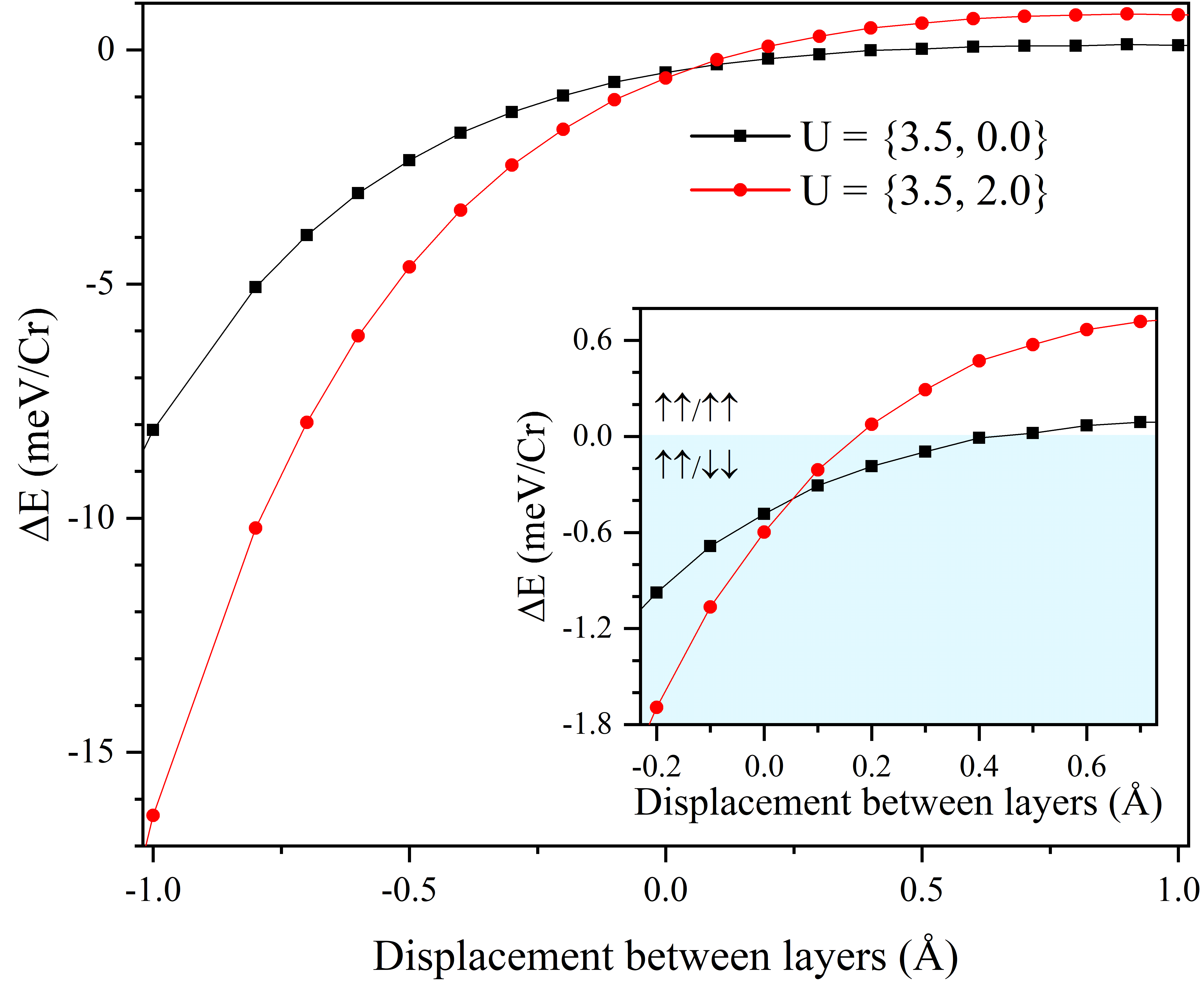}
    \caption{Bilayer energy difference $\Delta E$ ($\uparrow \uparrow / \downarrow \downarrow - \uparrow \uparrow / \uparrow \uparrow$) when the layer distance is changed. The equilibrium position is set at zero displacement.}
    \label{fig:distancia-biacapa}
\end{figure}

\end{document}